\documentclass[12pt, reqno]{amsart}

\usepackage[margin=.9in, a4paper, foot=.3in]{geometry}

\usepackage{graphicx, color}

\definecolor{darkblue}{rgb}{0,0,.5}
\definecolor{darkred}{rgb}{.5,0,0}
\definecolor{darkgreen}{rgb}{0,0.5,0}

\usepackage{hyperref}
\hypersetup {
    pdfstartview=FitH,
    colorlinks,
    citecolor=darkgreen,
    linkcolor=darkred,
    urlcolor=darkblue
}

\let\oldtocsection=\tocsection
\let\oldtocsubsection=\tocsubsection
\let\oldtocsubsubsection=\tocsubsubsection

\renewcommand{\tocsection}[2]{\hspace{0em}\oldtocsection{#1}{#2}}
\renewcommand{\tocsubsection}[2]{\hspace{1em}\oldtocsubsection{#1}{#2}}
\renewcommand{\tocsubsubsection}[2]{\hspace{2em}\oldtocsubsubsection{#1}{#2}}

\usepackage{float}

\usepackage[font=small]{caption}

\usepackage{cite}

\usepackage[noBBpl,slantedGreek]{mathpazo}
\usepackage[mathcal]{euscript}

\usepackage{amssymb}
\allowdisplaybreaks
\numberwithin{equation}{section}
\numberwithin{figure}{section}

\usepackage{mathtools}

\usepackage{ifthen}

\newcommand {\svee}[2][\null]{#2%
  \ifthenelse{\equal{#1}{\null}}{\,}{}%
  \check{{}_{#1}}%
  \ifthenelse{\equal{#1}{\null}}{\,}{}%
}

\newcommand {\interval}[2]{[#1 \, . \, . \, #2]}

\newcommand {\id}{\mathrm{id}}
\newcommand {\rme}{\mathrm e}
\newcommand {\rmS}{\mathrm S}

\newcommand {\bbC}{\mathbb C}
\newcommand {\bbZ}{\mathbb Z}

\newcommand {\calL}{\mathcal L}
\newcommand {\calR}{\mathcal R}
\newcommand {\calW}{\mathcal W}

\newcommand {\gothh}{\mathfrak h}
\newcommand {\gothg}{\mathfrak g}

\newcommand {\tgothh}{\widetilde{\mathfrak h}}
\newcommand {\tlgothg}{\widetilde{\mathcal L}(\mathfrak g)}
\newcommand {\uqglii}{\mathrm U_q(\mathfrak{gl}_2)}
\newcommand {\uqlg}{\mathrm U_q(\mathcal L(\mathfrak g))}
\newcommand {\uqhlslii}{\mathrm U_q(\widehat{\mathcal L}(\mathfrak{sl}_2))}
\newcommand {\uqlslii}{\mathrm U_q(\mathcal L(\mathfrak{sl}_2))}
\newcommand {\uqlsllpo}{\mathrm U_q(\mathcal L(\mathfrak{sl}_{l + 1}))}
\newcommand {\uqgllpo}{\mathrm U_q(\mathfrak{gl}_{l + 1})}

\DeclareMathOperator {\Aut}{Aut}
\DeclareMathOperator {\diag}{diag}
\DeclareMathOperator {\tr}{tr}

\title{Reduced qKZ equation and genuine qKZ equation}

\author[A. V. Razumov]{Alexander V. Razumov}
\address{Institute for High Energy Physics, NRC ``Kurchatov Institute", 142281 Protvino, Mos\-cow region, Russia}
\email{Alexander.Razumov@ihep.ru}

\begin{document}

\begin{abstract}
The work is devoted to the study of quantum integrable systems associated with quantum loop algebras. The recently obtained equation for the zero temperature inhomogeneous reduced density operator is analyzed. It is demonstrated that any solution of the corresponding qKZ equation generates a solution to this equation.
\end{abstract}

\maketitle

\tableofcontents

\section{Introduction}

In the paper \cite{KluNirRaz19a} a difference-type functional equation for the zero temperature inhomogeneous reduced density operator of a quantum integrable spin chain, called the reduced quantum Knizhnik--Zamolodchikov (qKZ) equation, was derived. The method based on the notion of a quantum group introduced by Drinfeld \cite{Dri87} and Jimbo \cite{Jim85} was used. In fact, the quantum integrable systems related to a special class of quantum groups, called the quantum loop algebras, were considered. A quantum loop algebra $\uqlg$ is defined for an arbitrary finite dimensional complex simple Lie algebra $\gothg$, and any representation of $\uqlg$ can be used to define a quantum integrable system.

The quantum group approach to the investigation of quantum integrable systems is based on the notion of the universal $R$-matrix being an element of the tensor product of two copies of the quantum loop algebra. The integrability objects are constructed by choosing representations for the factors of that tensor product. The properties of the integrability objects follow from the properties of the corresponding quantum loop algebra and its representations. A gentle introduction to integrability objects and their basic properties can be found in the papers \cite{BooGoeKluNirRaz13, BooGoeKluNirRaz14a}.

For the first time the quantum group approach was consistently used for constructing integrability objects and proving their properties by Bazhanov, Lukyanov and Zamolodchikov \cite{BazLukZam96, BazLukZam97, BazLukZam99}. They studied the quantum version of the KdV theory. Earlier and then the method proved to be efficient for studying other quantum integrable models. The integrability objects, such as $R$-operators, monodromy operators and $L$-operators were constructed \cite{KhoTol92, LevSoiStu93, ZhaGou94, BraGouZhaDel94, BraGouZha95, BazTsu08, BooGoeKluNirRaz10, BooGoeKluNirRaz11,  Raz13}. The respective sets of functional relations were found and proved \cite{BazHibKho02, Koj08, BazTsu08, BooGoeKluNirRaz14b, NirRaz14}.

The name used for the equation derived in the paper \cite{KluNirRaz19a} is explained by the fact, see the book by Jimbo and Miwa \cite{JimMiw95}, that in the case of the quantum loop algebra $\uqlslii$ the matrix elements of the inhomogeneous reduced density operator are directly related to the matrix elements of an appropriate product of vertex operators between certain representations of the quantum group $\uqhlslii$, which satisfy the qKZ equation \cite{FreRes92}. Thus, in this case any solution of the qKZ equation gives a solution of the reduced qKZ equation. In this paper, which can be considered as a continuation of \cite{KluNirRaz19a}, we show that this is so in the general case as well.

The plan of the paper is as follows. In section \ref{s:pois} we define quantum loop algebras, discuss the construction of $R$-operators and present the graphical depiction of their properties. The notion of the inhomogeneous reduced density operator is introduced in section~\ref{s:do}. The connection of the equations for the reduced inhomogeneous density operators with qKZ equations is derived in section \ref{s:rqearqe}. First, the case of the integrable systems related to the almost self-dual representations is discussed, and then the general case is considered. In the appendix we give the necessary information on the qKZ equations.

We actively use the graphical approach developed in the paper \cite{NirRaz19}. This approach has already proved to be useful in deriving the reduced qKZ equation and has confirmed its capabilities in our consideration.

\section{Preliminaries on integrability objects} \label{s:pois}

\subsection{Quantum loop algebras}

Let $\gothg$ be a finite dimensional complex simple Lie algebra of rank $l$ \cite{Ser01, Hum80}, $\gothh$ a Cartan subalgebra of $\gothg$, and $\Delta$ the root system of $\gothg$ relative to $\gothh$. Fix a system of simple roots $\alpha_i$, $i \in \interval{1}{l}$. It is known that the corresponding coroots $h_i$ form a basis of $\gothh$.
Remind that the Cartan matrix $A = (a_{i j})_{i, j \in \interval{1}{l}}$ of $\gothg$, where
\begin{equation*}
a_{i j} = \langle \alpha_j, \, h_i \rangle,
\end{equation*}
is symmetrizable. It means that there exists a diagonal matrix $D = \diag(d_1, \, \ldots, d_l)$, where $d_i$ are positive integers, such that the matrix $D A$ is symmetric. Such a matrix $D$ is defined up to a nonzero scalar factor. We fix the normalization of $D$ assuming that the integers $d_i$ are relatively prime. Following Kac, we denote by $\calL(\gothg)$ the loop algebra of $\gothg$ and by $\tlgothg$ its standard central extension. It is natural to denote by $\tgothh$ the arising extension of $\gothh$. For details see the monograph by Kac\cite{Kac90} and, in the form adopted for our purposes, the paper \cite{NirRaz19}.

Let $\hbar$ be a nonzero complex number such that $q = \exp \hbar$ is not a root of unity. We assume that
\begin{equation*}
q^\nu = \exp (\hbar \nu)
\end{equation*}
for any $\nu \in \bbC$. It is common to define the $q$-deformation of a number $\nu \in \bbC$ as
\begin{equation*}
[\nu]_q = \frac{q^\nu - q^{-\nu}}{q - q^{-1}}.
\end{equation*}
The quantum loop algebra $\uqlg$ is a unital associative $\bbC$-algebra generated by the elements $e_i$, $f_i$, $i \in \interval{0}{l}$, and $q^x$, $x \in \tgothh$, subjected to the corresponding defining relations, see, for example, the paper \cite{NirRaz19}. The quantum loop algebra $\uqlg$ is a Hopf algebra with the proper comultiplication $\Delta$, antipode $S$, and counit $\varepsilon$.

Let $\Pi \colon a \otimes b \in \uqlg \otimes \uqlg \mapsto b \otimes a \in \uqlg \otimes \uqlg$ be the linear operator permuting the factors of the tensor product. It can be shown that there exists a unique element $\calR \in \uqlg \otimes \uqlg$ connecting the comultiplication $\Delta$ with the inverse comultiplication $\Delta' = \Pi \circ \Delta$ in the sense that
\begin{equation*}
\Delta'(a) = \calR \, \Delta(a) \, \calR^{-1}
\end{equation*}
for any $a \in \uqlg$. The element $\calR$ is called the universal $R$-matrix.\footnote{In fact, for a quantum loop algebra defined as a complex algebra, the universal $R$-matrix exists only in some restricted sense, see, for example, the paper \cite{Tan92}, and the corresponding discussion in the paper \cite{NirRaz19} for the case of $\uqlsllpo$.}

\subsection{Spectral parameter}

We introduce spectral parameters in the following way. Assume that the quantum loop algebra $\uqlg$ is $\bbZ$-graded,
\begin{equation*}
\uqlg = \bigoplus_{m \in \bbZ} \uqlg_m, \qquad \uqlg_m \, \uqlg_n \subset \uqlg_{m + n},
\end{equation*}
so that any element $a \in \uqlg$ can be uniquely represented as
\begin{equation*}
a = \sum_{m \in \bbZ} a_m, \qquad a_m \in \uqlg_m.
\end{equation*}
Given $\zeta \in \bbC^\times$, we define the grading automorphism $\Gamma_\zeta$ by the equation
\begin{equation*}
\Gamma_\zeta(a) = \sum_{m \in \bbZ} \zeta^m a_m.
\end{equation*}
Now, for any representation $\varphi$ of $\uqlg$ we define the corresponding family $\varphi_\zeta$ of representations as
\begin{equation*}
\varphi_\zeta = \varphi \circ \Gamma_\zeta.
\end{equation*}
If $V$ is the $\uqlg$-module corresponding to the representation $\varphi$, we denote by $V_\zeta$ the $\uqlg$-module corresponding to the representation $\varphi_\zeta$.

In the present paper we endow $\uqlg$ with a $\bbZ$-gradation assuming that
\begin{equation*}
q^x \in \uqlg_0, \qquad e_i \in \uqlg_{s_i}, \qquad f_i \in \uqlg_{-s_i},
\end{equation*}
where $s_i$ are arbitrary integers. We denote
\begin{equation*}
s = \sum_{i = 0}^l a_i s_i,
\end{equation*}
where $a_i$ are the Kac labels of the Dynkin diagram associated with the extended Cartan matrix of $\gothg$.

\subsection{\texorpdfstring{$R$-operators and unitarity relation}{R-operators and unitarity relation}}

Let $V_1$, $V_2$ be two $\uqlg$-modules, $\varphi_1$, $\varphi_2$ the corresponding representations of $\uqlg$.\footnote{In the present paper we consider only finite dimensional highest weight $\uqlg$-modules, see for definition the papers \cite{FreRes99, MukYou14}}. We define the $R$-operator $R_{V_1 | V_2}(\zeta_1 | \zeta_2)$ by the equation
\begin{equation*}
\rho_{V_1 | V_2}(\zeta_1 | \zeta_2) R_{V_1 | V_2}(\zeta_1 | \zeta_2) = (\varphi_{1 \zeta_1} \otimes \varphi_{2 \zeta_2}) (\calR),
\end{equation*}
where $\zeta_1$ and $\zeta_2$ are two spectral parameters, and $\rho_{V_1 | V_2}(\zeta_1 | \zeta_2)$ is a scalar normalization factor. We choose the normalization factor so that
\begin{equation*}
\rho_{V_1 | V_2}(\zeta_1 \nu | \zeta_2 \nu) = \rho_{V_1 | V_2}(\zeta_1 | \zeta_2)
\end{equation*}
for any $\nu \in \bbC^\times$.  In this case
\begin{equation*}
R_{V_1 | V_2}(\zeta_1 \nu | \zeta_2 \nu) = R_{V_1 | V_2}(\zeta_1 | \zeta_2),
\end{equation*}
see the paper \cite{BooGoeKluNirRaz14a}. This means that $R_{V_1 | V_2}(\zeta_1 | \zeta_2)$ depends only on the combination
\begin{equation*}
\zeta\strut_{1 2} = \zeta\strut_1 \zeta\strut^{-1}_2,
\end{equation*}
and one can use $R$-operators depending on only one spectral parameter. Below we always use the above choice of the normalization, however, for our purposes it is more convenient to consider $R$-operators as depending on two spectral parameters.

\begin{figure}[t!]
\begin{minipage}{0.45\textwidth}
\centering
\includegraphics{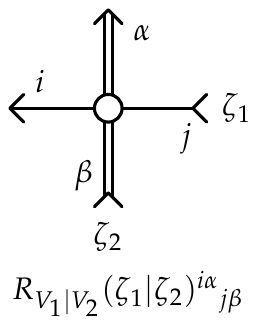}
\caption{}
\label{f:ro}
\end{minipage} \hfil
\begin{minipage}{0.45\textwidth}
\centering
\includegraphics{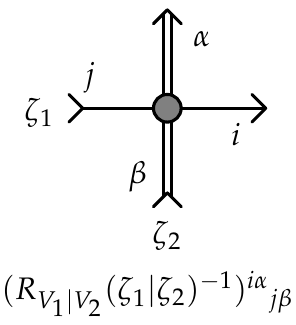}
\caption{}
\label{f:iro}
\end{minipage}
\end{figure}

For the matrix entries of the operators $R_{V_1 | V_2}(\zeta_1 | \zeta_2)$ and $R_{V_1 | V_2}(\zeta_1 | \zeta_2)^{-1}$ with respect to some bases of $V_1$ and $V_2$ we use the depiction given in figures \ref{f:ro} and \ref{f:iro}.

Let $\uqlg$-modules $V_1$ and $V_2$ be such that the $\uqlg \otimes \uqlg$-module $V_{1 \zeta_1} \otimes V_{2 \zeta_2}$ is simple for general values of the spectral parameters $\zeta_1$ and $\zeta_2$. In this case the following unitarity relation
\begin{equation*}
\check R_{V_1 | V_2}(\zeta_1 | \zeta_2) \check R_{V_2 | V_1}(\zeta_2 | \zeta_1) = C_{V_1 | V_2}(\zeta_1 | \zeta_2) \, \id_{V_2 \otimes V_1},
\end{equation*}
where $C_{V_1 | V_2}(\zeta_1 | \zeta_2)$ is some scalar function, is valid. Here and below we use the notation
\begin{equation*}
\check R_{V_1 | V_2}(\zeta_1 | \zeta_2) = P_{V_1 | V_2}  R_{V_1 | V_2}(\zeta_1 | \zeta_2)
\end{equation*}
with $P_{V_1 | V_2}$ being the permutation operator on the corresponding tensor product.

\begin{figure}[t!]
\centering
\includegraphics{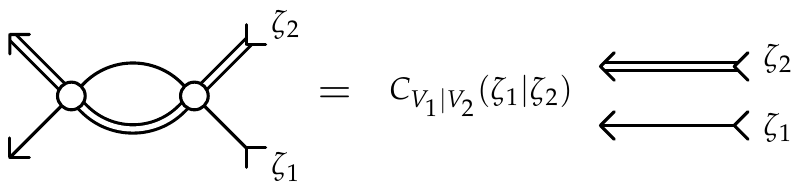}
\caption{}
\label{f:ura}
\end{figure}

\begin{figure}[t!]
\centering
\includegraphics{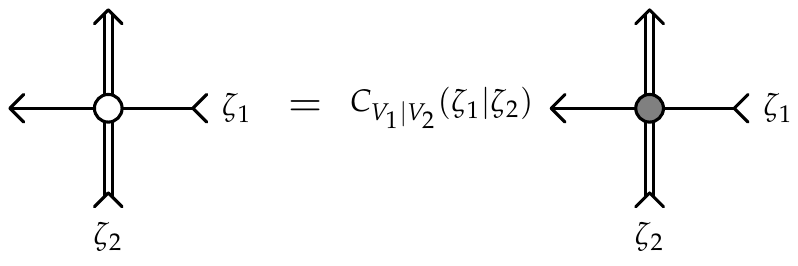}
\caption{}
\label{f:urb}
\end{figure}

It is convenient to use two different graphical forms of the unitarity relation given in figures \ref{f:ura} and \ref{f:urb}.

\subsection{Crossing relations}

For any $\uqlg$-module $V$ there are two dual modules usually denoted by $V^*$ and~${}^*V$. As vector spaces both modules coincide with the dual space $V^*$.\footnote{In the case of an infinite dimensional module $V$ one uses the restricted dual space, see, for example, the paper  \cite{NirRaz19}} The representation corresponding to the module $V^*$ is
defined by the equation
\begin{equation}
\varphi^*(a) = \varphi(S(a))^t, \label{fsa}
\end{equation}
and corresponding to the module ${}^*V$ by the equation
\begin{equation*}
{}^* \! \varphi = \varphi(S^{-1}(a))^t.
\end{equation*}
The modules $V^*$ and ${}^*V$ are isomorphic and we use only the first one. For this module we use the dotted variant of the line used for the module $V$.

\begin{figure}[t!]
\begin{minipage}{0.495\textwidth}
\centering
\includegraphics{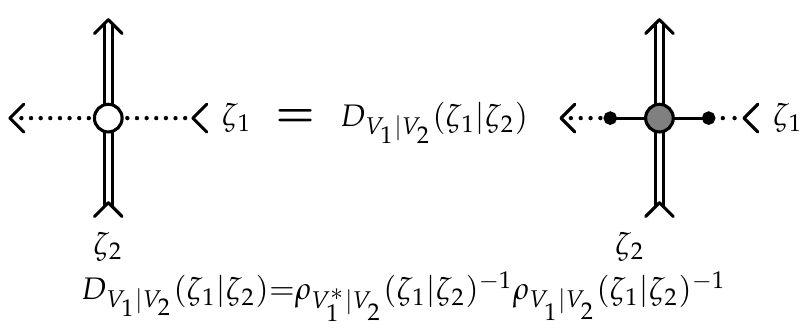}
\caption{}
\label{f:ca}
\end{minipage} \hfil
\begin{minipage}{0.495\textwidth}
\centering
\includegraphics{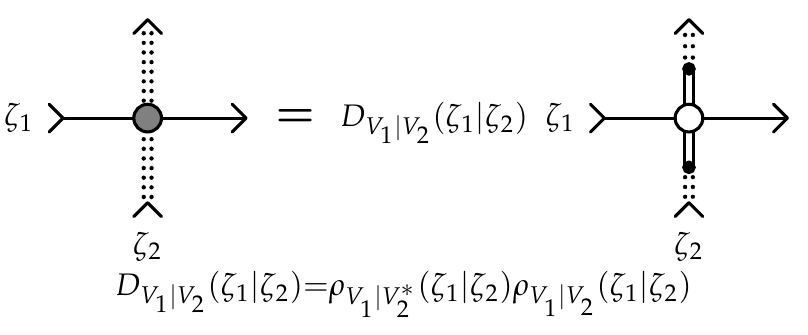}
\caption{}
\label{f:cb}
\end{minipage}
\end{figure}

\begin{figure}[t!]
\centering
\includegraphics{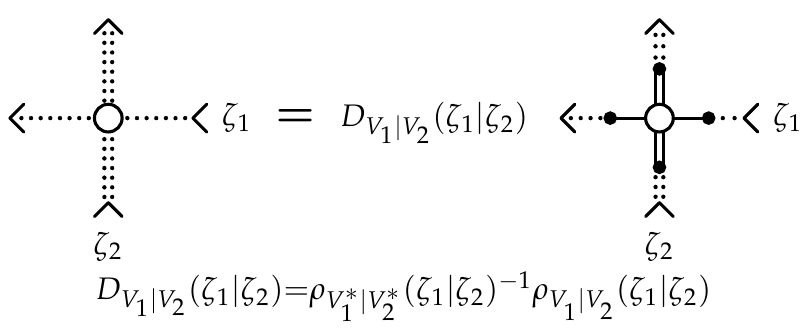}
\caption{}
\label{f:cc}
\end{figure}

By a crossing relation we mean any relation connecting an $R$-operator $R_{V_1 | V_2}(\zeta_1 | \zeta_2)$ with an $R$-operator for which one of the modules $V_1$ and $V_2$, or both of them, is replaced by a dual module, or consequences of such relations. The simplest crossing relations are depicted in figures \ref{f:ca}, \ref{f:cb} and \ref{f:cc}. Here and below fat dots mean the following change of the line type. The line corresponding to a module becomes the line corresponding to the dual module and vice versa, with the appropriate change of the direction.

\begin{figure}[t!]
\begin{minipage}{0.45\textwidth}
\centering
\includegraphics{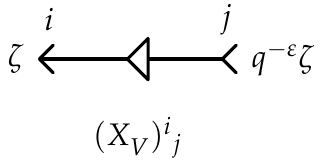}
\caption{}
\label{f:ox}
\end{minipage} \hfil
\begin{minipage}{0.45\textwidth}
\centering
\includegraphics{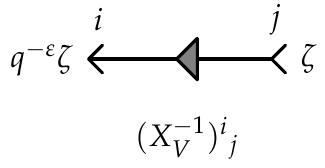}
\caption{}
\label{f:iox}
\end{minipage}
\end{figure}

\begin{figure}[t!]
\begin{minipage}{0.45\textwidth}
\centering
\includegraphics{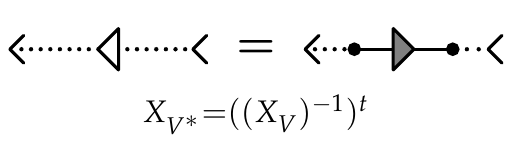}
\caption{}
\label{f:dox}
\end{minipage} \hfil
\begin{minipage}{0.45\textwidth}
\centering
\includegraphics{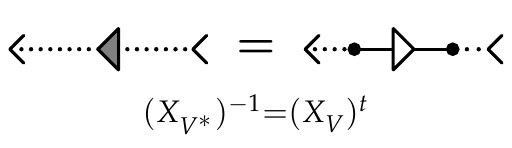}
\caption{}
\label{f:idox}
\end{minipage}
\end{figure}

The double dual representation $V^{**}_\zeta$ is isomorphic to $V_\zeta$ up to a redefinition of the spectral parameter. To be more concrete, we introduce the following element  
\begin{equation*}
x = - \sum_{i, j = 1}^l (2 d_i - (\theta | \theta) \svee{h} s_i/s) \, b_{i j} \, h_j
\end{equation*}
of $\tgothh$, see the paper \cite{NirRaz19}. Here $b_{i j}$, $i, j \in \interval{1}{l}$, are the matrix elements of
the matrix $B$ inverse to the Cartan matrix $A$ of $\gothg$, $\theta$ the highest root of $\gothg$, $\svee{h}$ the dual Coxeter number of $\gothg$, and $(\cdot | \cdot)$ denotes the invariant nondegenerate symmetric bilinear form on $\gothh$ normalized by the equation
\begin{equation*}
(\alpha_i | \alpha_i) = 2 d_i.
\end{equation*}
Now one can demonstrate that
\begin{equation*}
\varphi^{**}_\zeta(a) = \varphi(q^x) \, \varphi_{q^{-\varepsilon} \zeta}(a) \, \varphi(q^{-x})
\end{equation*}
for any $a \in \uqlg$. Here and afterwards we use the notation 
\begin{equation}
\varepsilon =(\theta | \theta) \svee{h}/s. \label{etts}
\end{equation}   
We denote
\begin{equation*}
X_V = \varphi(q^x)
\end{equation*}
and introduce for the matrix entries of $X_V$ and its inverse the graphical depiction given in figures \ref{f:ox} and \ref{f:iox}. The relation of the operators $X_{V^*}$ and $(X_{V^*})^{-1}$ with the operators $X_V$ and $(X_V)^{-1}$ is described by figures \ref{f:dox} and \ref{f:idox}.

\begin{figure}[t!]
\begin{minipage}{0.495\textwidth}
\centering
\includegraphics{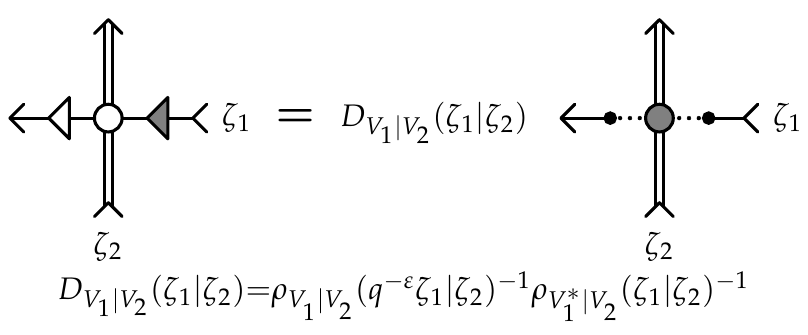}
\caption{}
\label{f:cd}
\end{minipage} \hfil
\begin{minipage}{0.495\textwidth}
\centering
\includegraphics{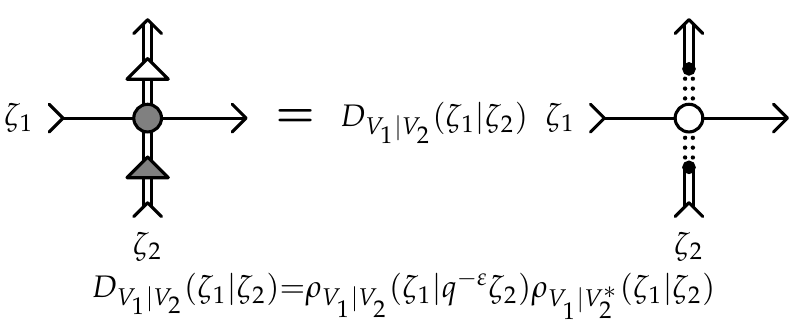}
\caption{}
\label{f:ce}
\end{minipage}
\end{figure}

\begin{figure}[t!]
\centering
\includegraphics{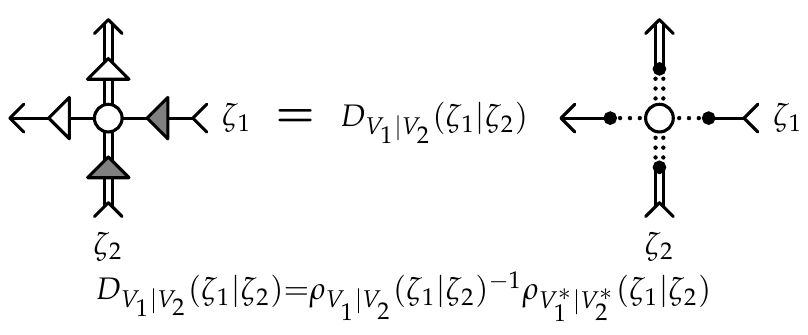}
\caption{}
\label{f:cf}
\end{figure}

\begin{figure}[t!]
\centering
\includegraphics{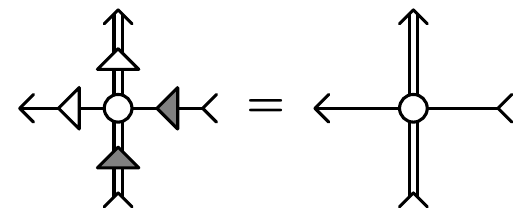}
\caption{}
\label{f:xi}
\end{figure}

The above described isomorphism leads to the crossing relations given in figures \ref{f:cd} and \ref{f:ce}.
Combining these relations we come to the crossing relation depicted in figure~\ref{f:cf}.
Now, using the crossing relation \ref{f:cc}, we obtain the invariance relation represented by figure \ref{f:xi}.
In fact, this relation follows from the fact that $q^x$ is a group-like element of $\uqlg$, see, for example, the paper~\cite{NirRaz19}.

More crossing relations and the corresponding proofs can be found in the review paper~\cite{NirRaz19}.

\subsection{Normalization}

Let $V_1$ and $V_2$ be $\uqlg$-modules, such that the module $V_{1 \zeta_1} \otimes V_{2 \zeta_2}$ is simple for general values of the spectral parameters $\zeta_1$ and $\zeta_2$. One can choose the normalization factor $\rho_{V_1 | V_2}(\zeta_1 | \zeta_2)$ so that the matrix elements of $R_{V_1 | V_2}(\zeta_1 | \zeta_2)$ are rational functions of the spectral parameters, and $R_{V_1 | V_2}(\zeta_1 | \zeta_2)$ satisfies the unitarity relation
\begin{equation}
\check R_{V_1 | V_2}(\zeta_1 | \zeta_2) \check R_{V_2 | V_1}(\zeta_2 | \zeta_1) = \id_{V_2 \otimes V_1}, \label{ur}
\end{equation}
see \cite[Propositions 9.5.3 and 9.5.5]{EtiFreKir98}. Denote this normalization factor as $\rho^0_{V_1 | V_2}(\zeta_1 | \zeta_2)$ and the corresponding $R$-operator as $R^0_{V_1 | V_2} (\zeta_1 | \zeta_2)$. The normalization in question is defined by the relation
\begin{equation}
R^0_{V_1 | V_2}(\zeta_1 | \zeta_2) (v^{(1)}_0 \otimes v^{(2)}_0) = v^{(1)}_0 \otimes v^{(2)}_0, \label{rvv}
\end{equation}
where $v^{(1)}_0$ and $v^{(2)}_0$ are the highest weight vectors of the modules $V_1$ and $V_2$.

In this paper we consider the case when each of the modules $V_1$ and $V_2$ coincides either with a fixed module $V$ or with the dual module $V^*$. Hence, we have four $R$-operators, and relate all normalization factors to $\rho^0_{V | V}(\zeta_1 | \zeta_2)$ in the following way
\begin{gather*}
\rho^0_{V^* | V}(\zeta_1 | \zeta_2) = \rho^0_{V | V}(\zeta_1 | \zeta_2)^{-1}, \qquad \rho^0_{V | V^*}(\zeta_1 | \zeta_2) = \rho^0_{V | V}(\zeta_1 | \zeta_2)^{-1}, \\
\rho^0_{V^* | V^*}(\zeta_1 | \zeta_2) = \rho^0_{V | V}(\zeta_1 | \zeta_2).
\end{gather*}
We denote the corresponding $R$-operators as $R^0_{V^* | V}(\zeta_1 | \zeta_2)$, $R^0_{V | V^*}(\zeta_1 | \zeta_2)$ and $R^0_{V^* | V^*}(\zeta_1 | \zeta_2)$. These $R$-operators satisfy the unitarity relations of the form (\ref{ur}), see the paper \cite{KluNirRaz19a}. Moreover, the factors $D_{V_1 | V_2}(\zeta_1 | \zeta_2)$ entering the crossing relations represented by figures \ref{f:ca}--\ref{f:cc} and \ref{f:cf} with $V_1 = V_2 = V$ become equal to $1$.

The general normalization factor of $R_{V | V}(\zeta_1 | \zeta_2)$ leading to the unitarity relation (\ref{ur}) for $V_1 = V_2 = V$ is
\begin{equation}
\rho_{V | V}(\zeta_1 | \zeta_2) = \kappa_{V | V}(\zeta_1 | \zeta_2) \, \rho^0_{V | V}(\zeta_1 | \zeta_2), \label{rkr}
\end{equation}
where the function $\kappa_{V | V}(\zeta_1 | \zeta_2)$ satisfies the equation
\begin{equation}
\kappa_{V | V}(\zeta_1 | \zeta_2) \, \kappa_{V | V}(\zeta_2 | \zeta_1) = 1. \label{kvvkvv}
\end{equation}
This function is fixed below. Using this general normalization factor, we adjust the normalizations factors of the $R$-operators under consideration to have
\begin{gather*}
\rho_{V^* | V}(\zeta_1 | \zeta_2) = \rho_{V | V}(\zeta_1 | \zeta_2)^{-1}, \qquad \rho_{V | V^*}(\zeta_1 | \zeta_2) = \rho_{V | V}(\zeta_1 | \zeta_2)^{-1}, \\
\rho_{V^* | V^*}(\zeta_1 | \zeta_2) = \rho_{V | V}(\zeta_1 | \zeta_2).
\end{gather*}
Here the factors $D_{V_1 | V_2}(\zeta_1 | \zeta_2)$ entering the crossing relations represented by figures \ref{f:ca}--\ref{f:cc} and \ref{f:cf} with $V_1 = V_2 = V$, remain equal to $1$.

The described above normalization was used in the derivation of the reduced qKZ equation \cite{KluNirRaz19a}. In addition, it was assumed there that the module $V$ is such that the initial condition
\begin{equation*}
R^0_{V | V}(1 | 1) = c_V P_{V | V}
\end{equation*}
is valid for some nonzero constant $c_V$. In this paper, we also use this assumption. Possibly changing the sign of $R^0_{V | V}(\zeta_1 | \zeta_2)$, we make $c_V$ equal to 1 without destroying the form of the unitarity and crossing relations. In order to preserve the initial condition in the general case, we assume that the function $\kappa_{V | V}(\zeta_1 | \zeta_2)$ satisfies the additional requirement
\begin{equation}
\kappa_{V | V}(1 | 1) = 1. \label{kvvo}
\end{equation}

\section{Density operator} \label{s:do}

\subsection{Inhomogeneous density operator}

The density operator $\Psi$ of a quantum statistical system in equilibrium with a reservoir at the absolute temperature $T$ is given by the equation 
\begin{equation*}
\Psi = \frac{1}{Z} \, \rme^{- \beta H}, \qquad \beta = \frac{1}{k_{\mathrm B} T},
\end{equation*}
where $k_B$ is the Boltzmann constant, $H$ is the Hamiltonian, and $Z$ the partition function of the system defined as
\begin{equation*}
Z = \tr \rme^{- \beta H}.
\end{equation*}
The expectation value of an arbitrary observable $F$ is
\begin{equation*}
\langle F \rangle = \tr (F \Psi) = \frac{1}{Z} \tr (F  \, \rme^{- \beta H}).
\end{equation*}
In particular, in this way one obtains correlators and correlation functions of the system.

In the present paper we consider the integrable spin chains associated with quantum loop groups $\uqlg$. As usually, the investigation of the problem starts with a spin chain of finite length $L$. The thermodynamic limit would be obtained when $L \to \infty$. However, the existence of a sensible limit over $L$ is doubtful. Therefore, we proceed to the density operator which allows to find expectation values only for local observables which are nontrivial only on a segment of a finite length $n$. We call this density operator the reduced density operator and denote it by $\Psi_n$.

\begin{figure}[t!]
\begin{minipage}{0.45\textwidth}
\centering
\includegraphics{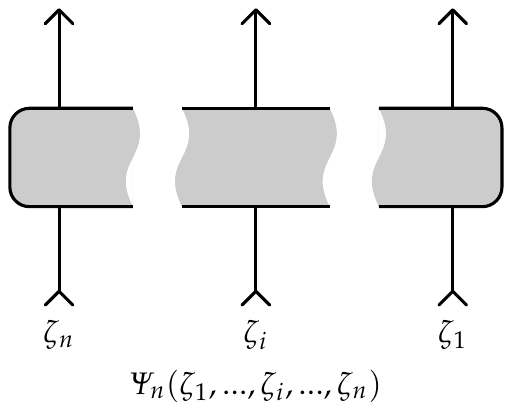}
\caption{}
\label{f:doa}
\end{minipage} \hfil
\begin{minipage}{0.45\textwidth}
\centering
\includegraphics{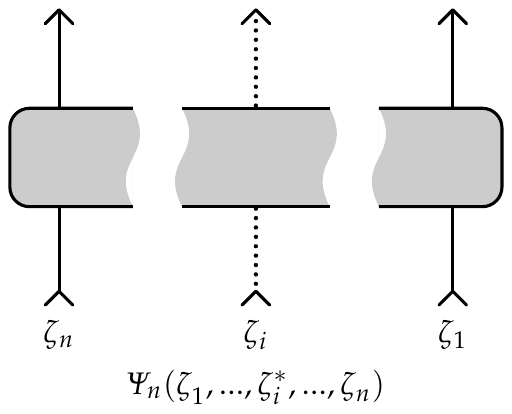}
\caption{}
\label{f:dob}
\end{minipage}
\end{figure}

The reduced density operator acts on the state space $V^{\otimes n}$, where $V$ is some $\uqlg$-module. To find equations satisfied by $\Psi_n$ one considers the inhomogeneous reduced density operator. To this end one introduces spectral parameters $\zeta_1$, $\ldots$, $\zeta_n$ and treat the $i$th factor of $V^{\otimes n}$ as the module $V_{\zeta_i}$.
Now the reduced density operator depends on the spectral parameters and we denote it as $\Psi_n(\zeta_1, \, \ldots, \, \zeta_n)$ and use the graphical image given in figure \ref{f:doa}.
For details we refer to the paper \cite{KluNirRaz19a}. It is convenient to define an auxiliary reduced density operators acting on the space $V^{\otimes n}$ where one of the factors is replaced by the dual module $V^*$. We denote such reduced density operator as $\Psi_n(\zeta^{}_1, \, \ldots, \, \zeta^*_i, \, \ldots, \, \zeta^{}_n)$, having in mind that $\zeta^*_i$ it is actually $\zeta^{}_i$ but associated with the dual module. The graphical image of $\Psi_n(\zeta^{}_1, \, \ldots, \, \zeta^*_i, \, \ldots, \, \zeta^{}_n)$ can be seen in figure \ref{f:dob}.

\begin{figure}[t!]
\centering
\begin{minipage}{0.4\textwidth}
\centering
\includegraphics{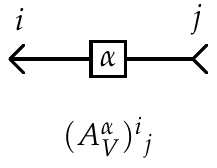}
\caption{}
\label{f:oa}
\end{minipage} \hfil
\begin{minipage}{0.4\textwidth}
\centering
\includegraphics{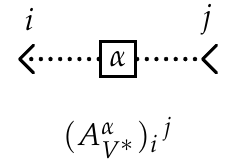}
\caption{}
\label{f:dao}
\end{minipage}
\end{figure}

We introduce also some twisting. In the case of $\uqlslii$, twisting is introduced to construct the fermionic operators \cite{BooJimMiwSmiTak07, BooJimMiwSmiTak09}. In general case a construction of the fermionic operators is not known yet. However, we introduce twisting for possible future needs. In the framework of the quantum group approach twisting is
defined by a choice of a group-like element of the quantum loop algebra. We choose for this goal the following element 
\begin{equation*}
a = q^{\sum_{i = 1}^l \alpha_i h_i}
\end{equation*}
where $\alpha_i$ are arbitrary complex numbers. We denote
\begin{equation*}
A^\alpha_V = \varphi(a), \qquad A^\alpha_{V^*} = \varphi^*(a)
\end{equation*}
and use for the matrix elements of the operators $A^\nu_V$ and $A^\nu_{V^*}$ the depiction given in figures \ref{f:oa} and \ref{f:dao}. By construction, we have
\begin{equation}
(A_{V_1}(\zeta_1) \otimes A_{V_2}(\zeta_2)) R_{V_1 | V_2}(\zeta_1 | \zeta_2) = R_{V_1 | V_2}(\zeta_1 | \zeta_2) (A_{V_1}(\zeta_1) \otimes A_{V_2}(\zeta_2)), \label{aarraa}
\end{equation}
where each of $V_1$ and $V_2$ is either $V$ or $V^*$ \cite{NirRaz19}.

\subsection{Reduced qKZ equation}

\begin{figure}[t!]
\centering
\includegraphics{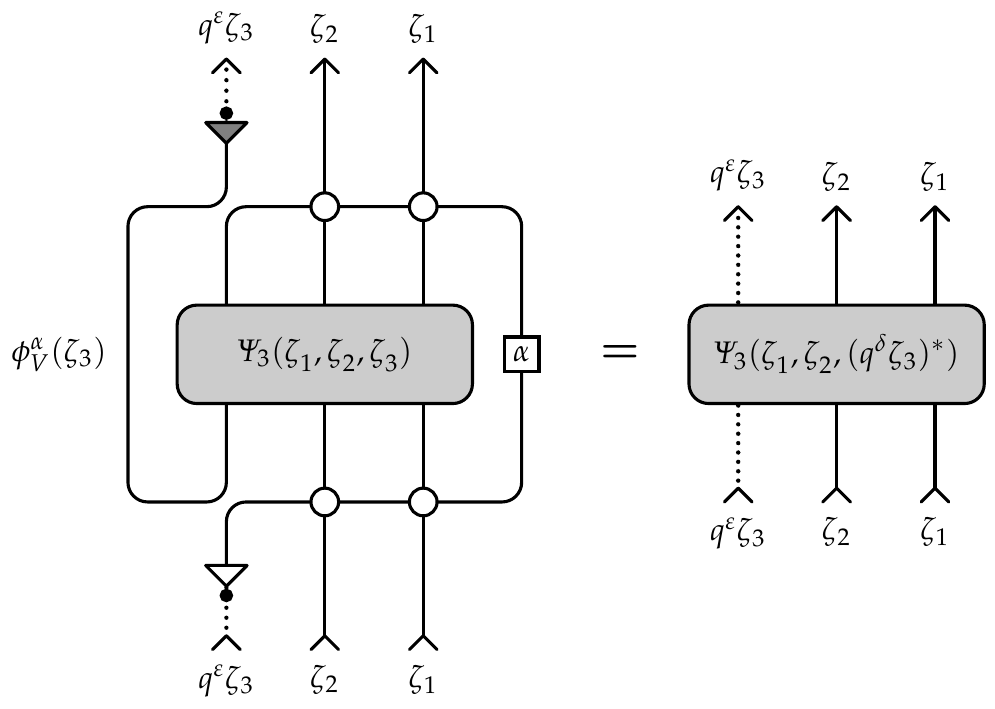}
\caption{}
\label{f:fe}
\end{figure}

\begin{figure}[t!]
\centering
\includegraphics{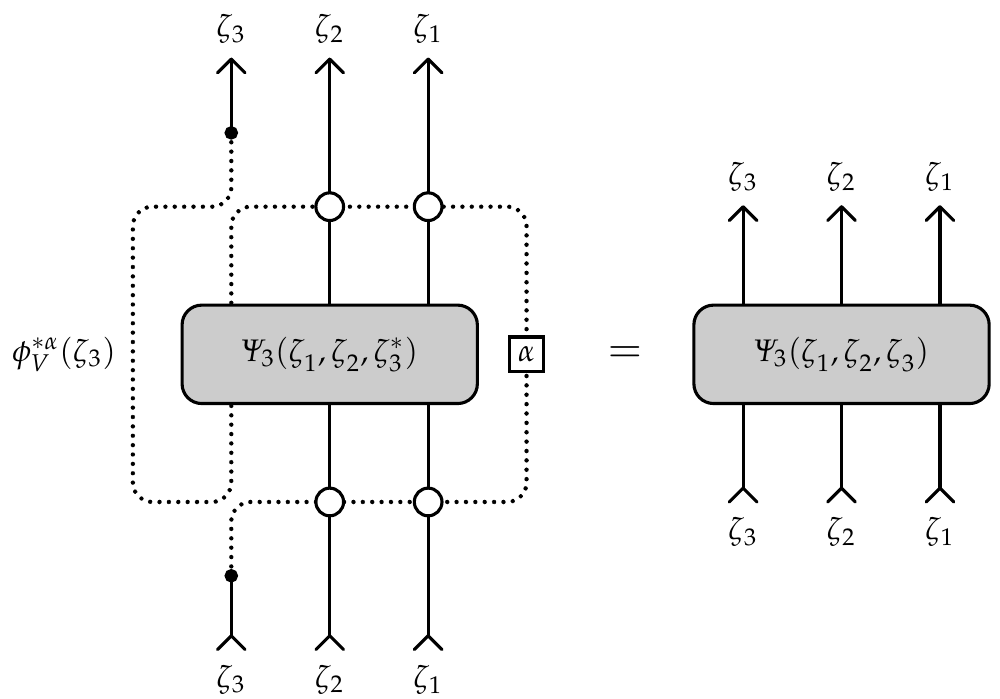}
\caption{}
\label{f:se}
\end{figure}

\begin{figure}[t!]
\centering
\includegraphics{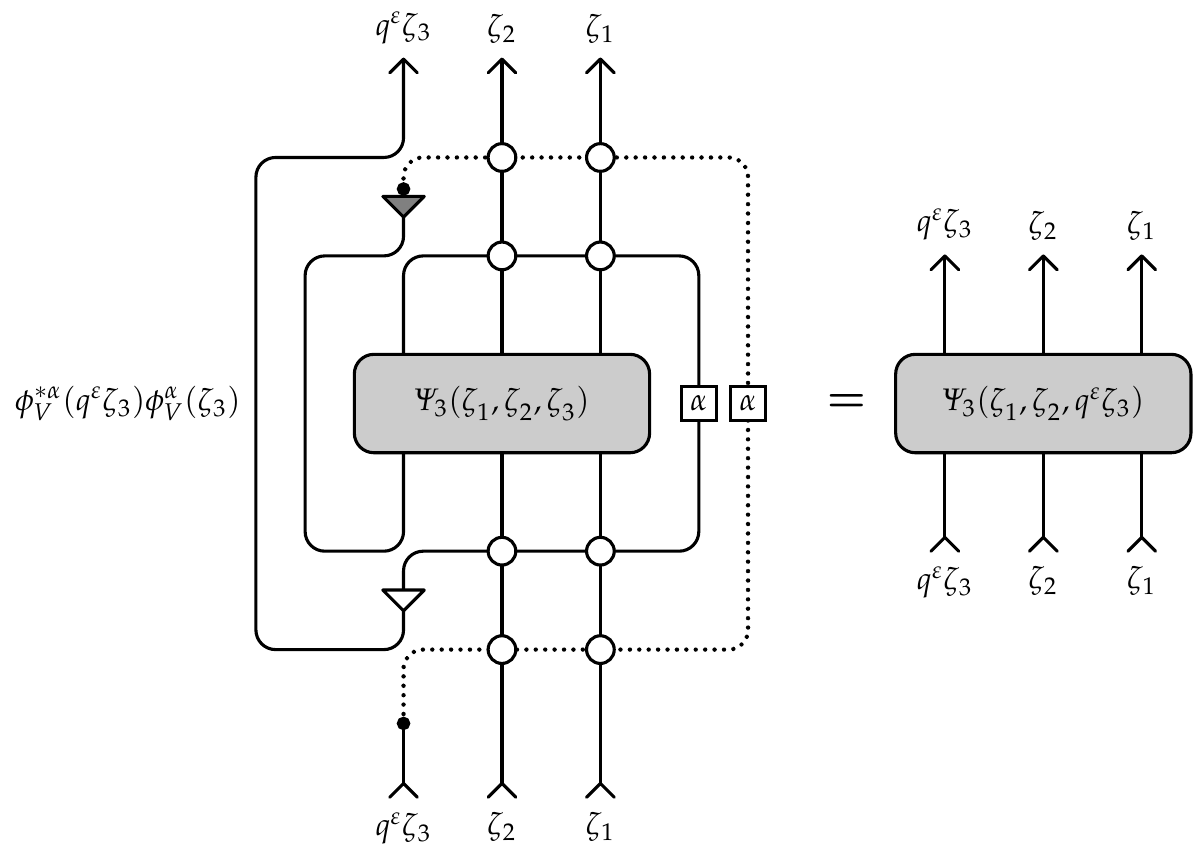}
\caption{}
\label{f:fine}
\end{figure}

It is shown in the paper \cite{KluNirRaz19a} that the zero-temperature limit of the operator $\Psi_n(\zeta_1, \ldots, \zeta_n)$ satisfies some difference equation. In fact, it is convenient to represent it as a system of two equations. The graphical image of the first equation is given in figure \ref{f:fe}.
Here and below we give graphical images for the case where $n = 3$. It is enough to understand the case of an arbitrary $n$. For the definition of the function $\phi^\alpha_V(\zeta)$ we refer to the paper \cite{KluNirRaz19a}. It is worth to note that in the case where $\alpha_i = 0$ for any $i \in \interval{1}{l}$ we have $\phi^\alpha_V(\zeta) = 1$. The graphical image of the second equation is given in figure \ref{f:se}.
For the definition of the function $\phi^{* \alpha}_V(\zeta)$ we again refer to the paper \cite{KluNirRaz19a}. Here as before in the case where $\alpha_i = 0$ for any $i \in \interval{1}{l}$ we have $\phi^{* \alpha}_V(\zeta) = 1$. The graphical image of the final equation can be obtained by combining the graphical equations given in figures \ref{f:fe} and \ref{f:se}, see figure \ref{f:fine}.
We call this equation the reduced qKZ equation. This is due to the fact, see the book by Jimbo and Miwa \cite{JimMiw95}, that in the case of $\uqlslii$ the matrix elements of the inhomogeneous reduced density operator are directly related to the matrix elements of an appropriate product of vertex operators between certain representations of the quantum group $\uqhlslii$, which satisfy qKZ equation \cite{FreRes92}. Thus, in this case any solution of the qKZ equation gives a solution of the reduced qKZ equation. Below we show that this is so in the general case as well.

To conclude this section, we note that the density operator by construction satisfies the relations
\begin{multline}
\check R^{(i, \, i + 1)}_{V | V} (\zeta_i | \zeta_{i + 1}) \Psi_n(\zeta_1, \ldots, \zeta_i, \zeta_{i + 1}, \ldots, \zeta_n) \\
= \Psi_n(\zeta_1, \ldots, \zeta_{i + 1}, \zeta_i, \ldots, \zeta_n) \check R^{(i, \, i + 1)}_{V | V} (\zeta_i | \zeta_{i + 1}) \label{rpr} 
\end{multline}
for each $i \in \interval{1}{n - 1}$. We consider these relations as a part of the reduced qKZ equation.

\section{Reduced qKZ equation as reduction of qKZ equation} \label{s:rqearqe}

\subsection{Almost self-dual representations}

Let $\varphi$ be a representation of $\uqlg$ and $V$ the corresponding $\uqlg$-module. Assume that
\begin{equation*}
\varphi^*_\zeta(a) = O\strut_V \, \varphi_{q^\delta \zeta}(a) \, O\strut^{-1}_V,
\end{equation*}
for some $O_V \in \Aut(V)$ and $\delta \in \bbC$. It means that the dual representation $\varphi^*_\zeta$ is isomorphic to the representation $\varphi_\zeta$ up to a redefinition of the spectral parameter. In this case we say that we deal with an almost self-dual representation. Here the graphical equation given in figure \ref{f:fe} can be made a closed equation.

\begin{figure}[t!]
\centering
\begin{minipage}{0.4\textwidth}
\centering
\includegraphics{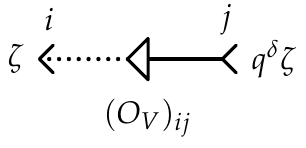}
\caption{}
\label{f:oo}
\end{minipage} \hfil
\begin{minipage}{0.4\textwidth}
\centering
\includegraphics{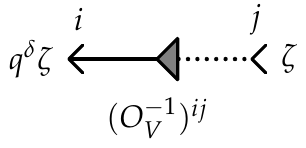}
\caption{}
\label{f:ioo}
\end{minipage}
\end{figure}

\begin{figure}[t!]
\centering
\begin{minipage}{0.49\textwidth}
\centering
\includegraphics{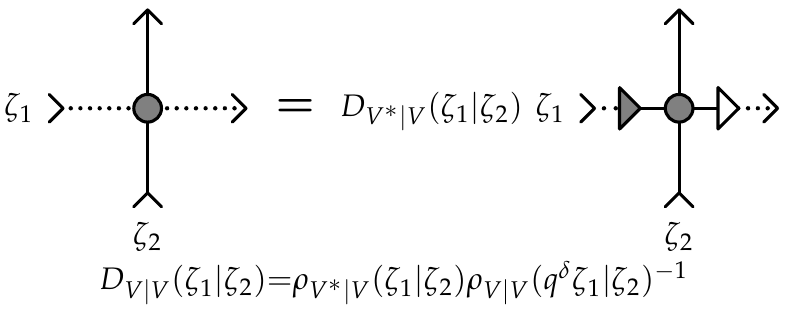}
\caption{}
\label{f:fda}
\end{minipage} \hfil
\begin{minipage}{0.49\textwidth}
\centering
\includegraphics{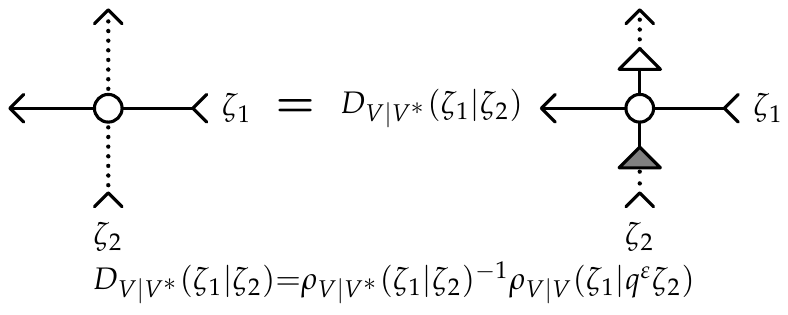}
\caption{}
\label{f:fdb}
\end{minipage}
\end{figure}

\begin{figure}[t!]
\centering
\begin{minipage}{0.49\textwidth}
\centering
\includegraphics{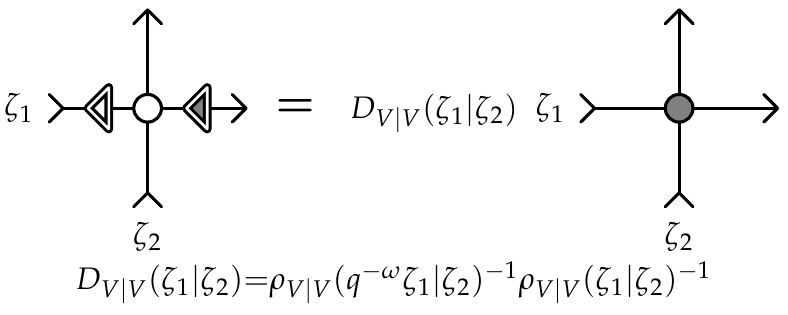}
\caption{}
\label{f:csl2a}
\end{minipage} \hfil
\begin{minipage}{0.49\textwidth}
\centering
\includegraphics{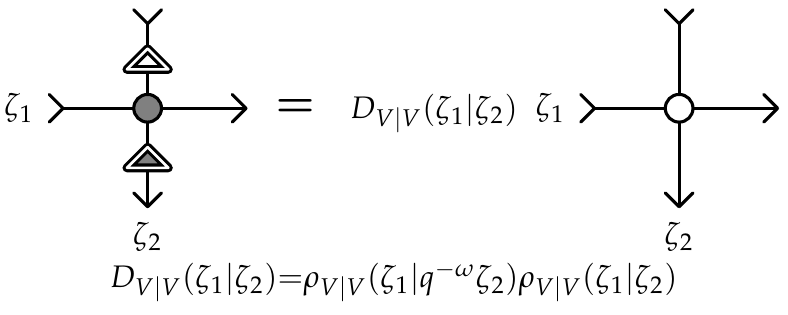}
\caption{}
\label{f:csl2b}
\end{minipage}
\end{figure}

\begin{figure}[t!]
\centering
\includegraphics{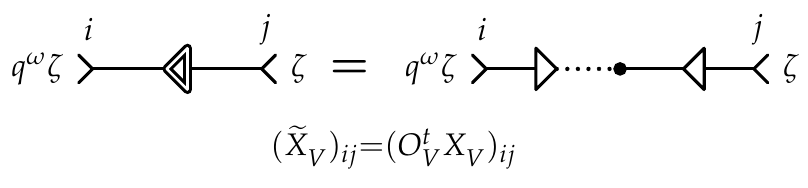}
\caption{}
\label{f:op}
\end{figure}

\begin{figure}[t!]
\centering
\includegraphics{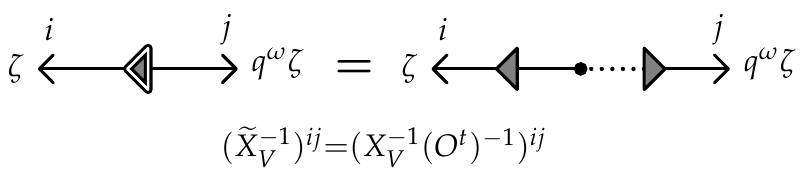}
\caption{}
\label{f:iop}
\end{figure}

\begin{figure}[t!]
\centering
\includegraphics{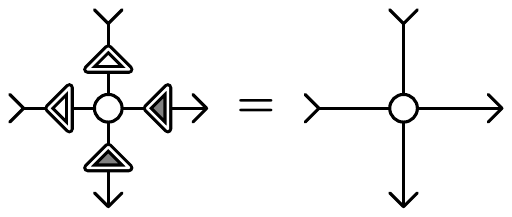}
\caption{}
\label{f:pi}
\end{figure}

The graphical depiction of the matrix entries of the operators $O^{}_V$ and $O_V^{-1}$ is given in figures \ref{f:oo} and \ref{f:ioo}.
Using these operators, we can find some relations between $R$-operators. Below we need two such relations depicted in figures \ref{f:fda} and \ref{f:fdb}.
Starting from the equation presented in figure \ref{f:cd} and using the equation given by figure \ref{f:fda}, we come to the equation depicted in figure \ref{f:csl2a}, where $\omega = \varepsilon + \delta$.
Here we introduce the operator
\begin{equation*}
\widetilde X^{}_V = O_V^t X^{}_V
\end{equation*}
and use for it and its inverse the graphical representation given in figures \ref{f:op} and \ref{f:iop}. In the same way, using figures \ref{f:fdb} and \ref{f:ce}, we come to the equation in figure \ref{f:csl2b}. The combination of the equations represented by figures \ref{f:csl2a} and \ref{f:csl2b} leads to the invariance relation which can be seen in figure \ref{f:pi}.

We use the normalization (\ref{rkr}), where $\rho^0_{V | V}(\zeta_1 | \zeta_2)$ is determined by the condition (\ref{rvv}). As the function $\kappa_{V | V}(\zeta_1 | \zeta_2)$ we take a solution of the difference equation
\begin{equation}
\rho^0_{V | V}(q^{-\omega} \zeta_1 | \zeta_2)^{-1} \rho^0_{V | V}(\zeta_1 | \zeta_2)^{-1} \kappa_{V | V}(q^{-\omega} \zeta_1 | \zeta_2)^{-1} \kappa_{V | V}(\zeta_1 | \zeta_2)^{-1}  = d_{V | V}. \label{rrkk}
\end{equation}
for some complex constant $d_{V | V}$. In this case the coefficients $D_{V | V}(\zeta_1 | \zeta_2)$ entering the crossing relations given in figures \ref{f:csl2a} and \ref{f:csl2b} turn into $d^{}_{V | V}$ and $d_{V | V}^{-1}$ respectively. We also assume that the conditions (\ref{kvvkvv}) and (\ref{kvvo}) are satisfied.

\begin{figure}[t!]
\centering
\includegraphics{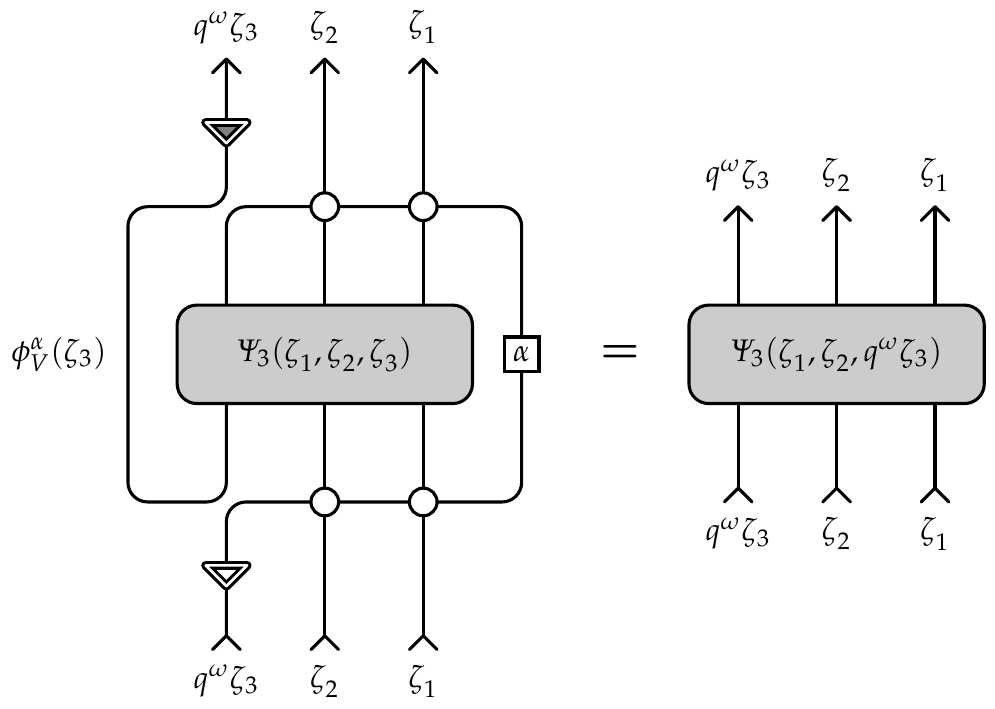}
\caption{}
\label{f:fesl2}
\end{figure}

By studying the paper \cite{KluNirRaz19a}, we conclude that the appearance of the operators $X\strut_V$ and $X_V^{-1}$ in the equation given by figure \ref{f:fe} is due to the fact that on the last step of its derivation one of the lines becomes going in the wrong direction. To reverse this line, one uses the crossing relations presented in figures \ref{f:ca} and \ref{f:cb}. This leads to appearance of the dual module and an auxiliary reduced density operator. In the case under consideration we can use for this goal the crossing relations given in figures \ref{f:csl2a} and \ref{f:csl2b} and come to a closed equation which is given in figure~\ref{f:fesl2}.

\begin{figure}[t!]
\centering
\includegraphics{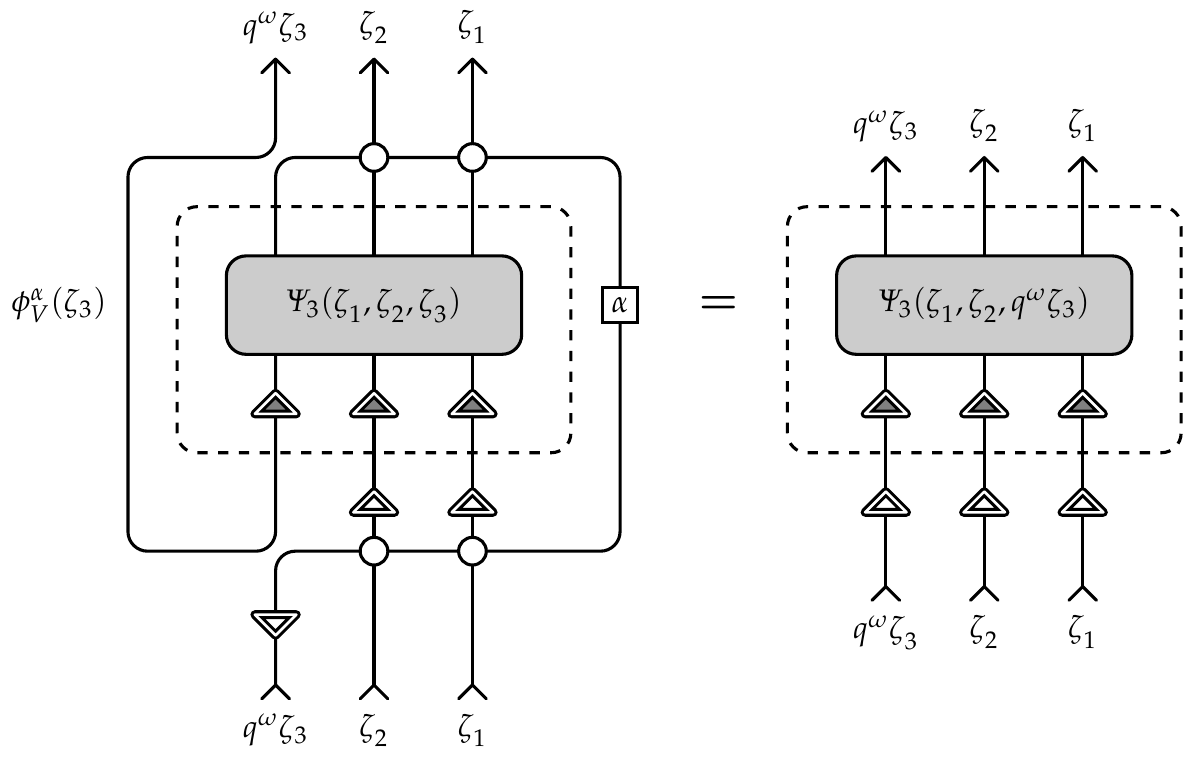}
\caption{}
\label{f:fesl2a}
\end{figure}

\begin{figure}
\centering
\includegraphics{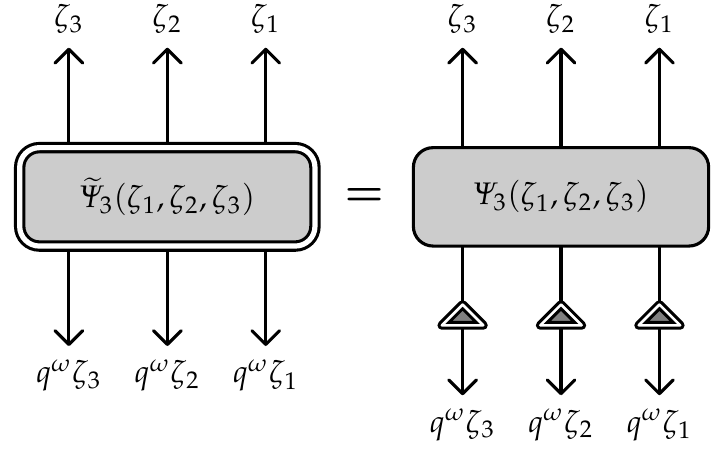}
\caption{}
\label{f:rdtd}
\end{figure}

\begin{figure}[t!]
\centering
\includegraphics{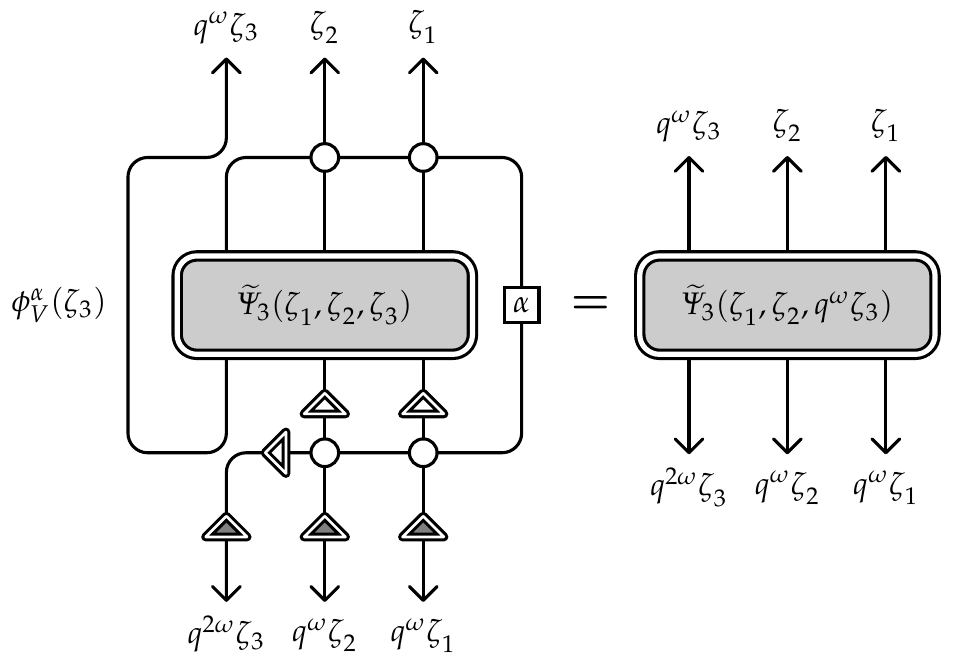}
\caption{}
\label{f:fesl2b}
\end{figure}

\begin{figure}[t!]
\centering
\includegraphics{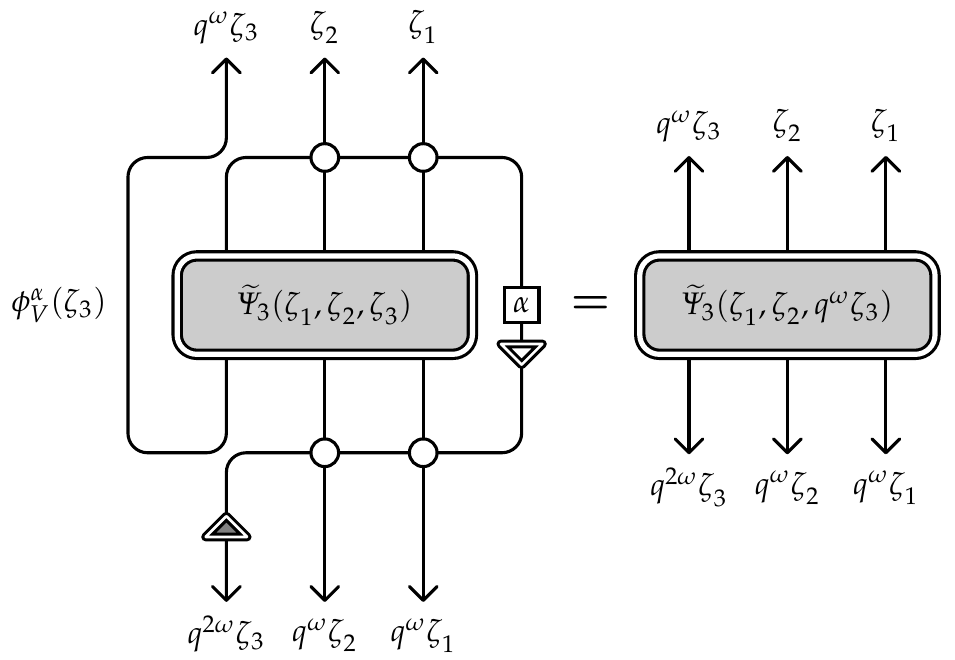}
\caption{}
\label{f:fesl2c}
\end{figure}

\begin{figure}[t!]
\centering
\includegraphics{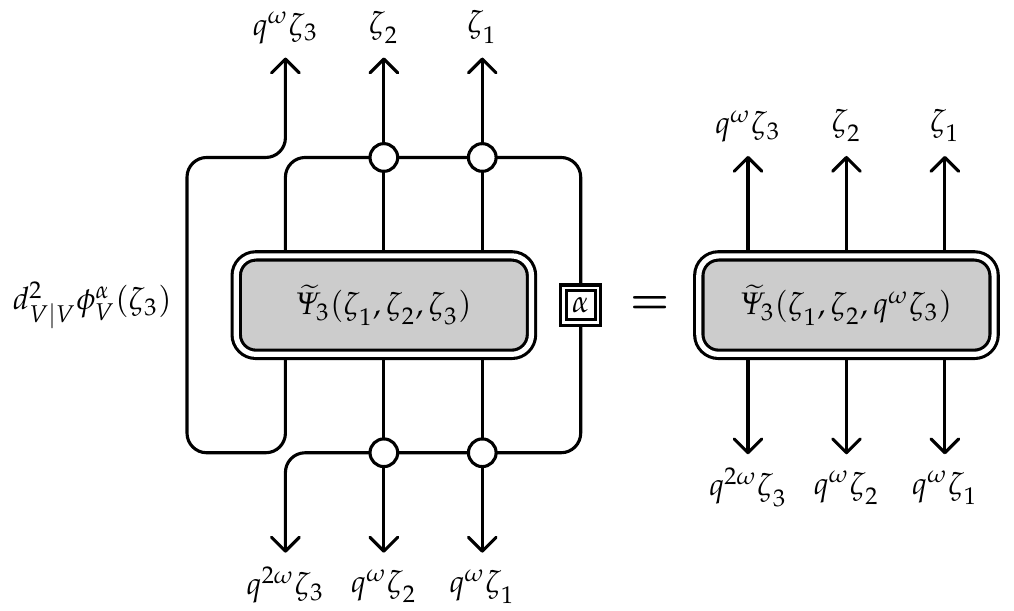}
\caption{}
\label{f:fesl2d}
\end{figure}

\begin{figure}[t!]
\centering
\includegraphics{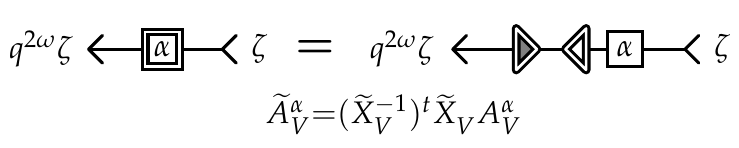}
\caption{}
\label{f:rtsl2}
\end{figure}

Redraw the obtained equation as is given in figure \ref{f:fesl2a}. Now it is natural to introduce the object $\widetilde \Psi_n(\zeta_1, \ldots, \zeta_n)$ defined by figure \ref{f:rdtd}.
Then the equation in figure \ref{f:fesl2a} takes the form of the equation in figure \ref{f:fesl2b}.
Using the invariance relation represented by figure~\ref{f:pi}, we move the leftmost white triangle to the right and come to the equation depicted by figure~\ref{f:fesl2c}.
Finally, with the help of the crossing relation \ref{f:csl2a} we move the leftmost grayed triangle to the right and obtain the graphical equation presented in figure \ref{f:fesl2d},
where the modified twisting operator $\widetilde A^\alpha_V$ is described by figure \ref{f:rtsl2}.

\begin{figure}[t!]
\centering
\includegraphics{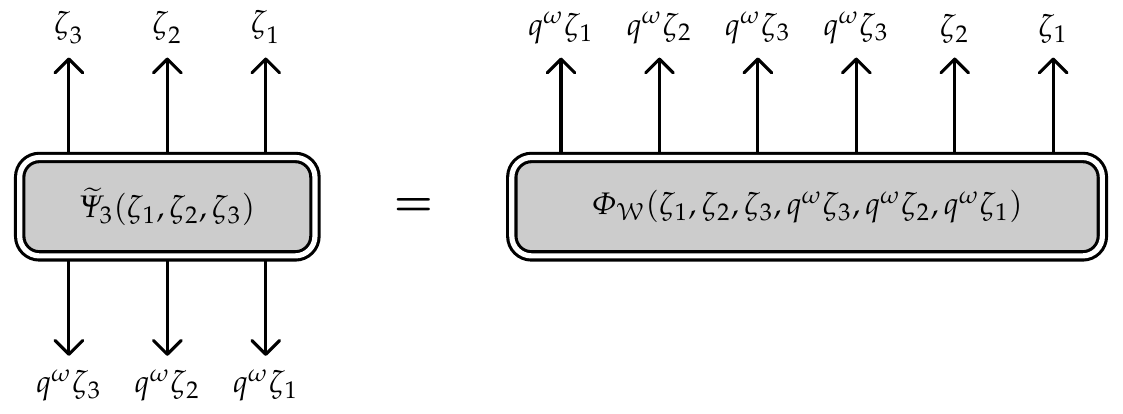}
\caption{}
\label{f:asl2}
\end{figure}

\begin{figure}[t!]
\centering
\includegraphics{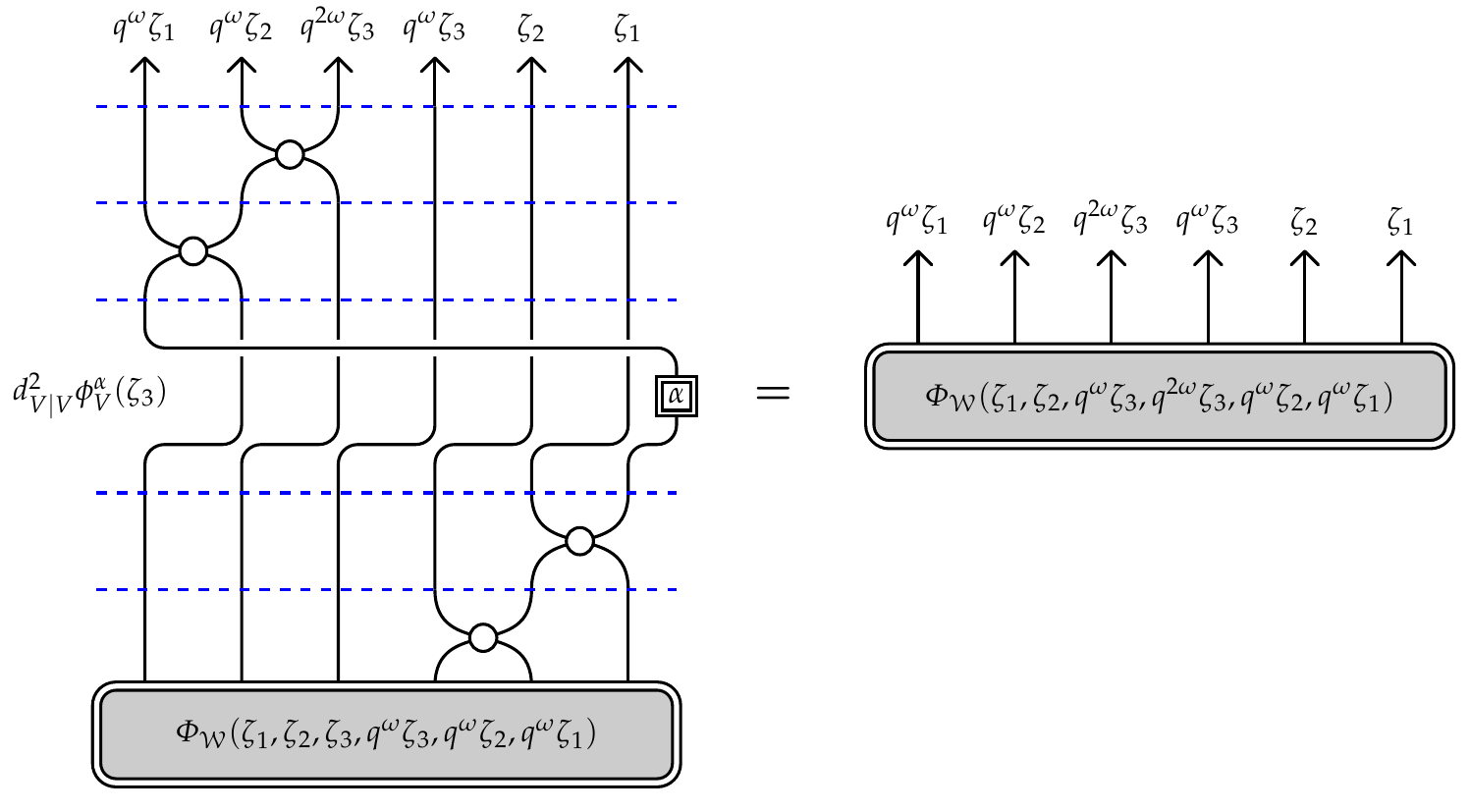}
\caption{}
\label{f:rrqkzsl2}
\end{figure}

One can try to construct a solution to this equation directly as it was done in the paper~\cite{BooJimMiwSmiTak06b} for the case of the representation of $\uqlslii$ generated from the representation $\pi^{(1/2, \, -1/2)}$ of $\uqglii$ by the Jimbo's homomorphism \cite{Jim86a}, see also the paper \cite{NirRaz19} for the relevant definitions. Another way is as follows. Let $N = 2n$, and $\Phi_\calW$ be the mapping described in the appendix, for the case where $W_i = V$ for all $i \in \interval{1}{N}$. Consider the ansatz
\begin{equation*}
\widetilde \Psi_n(\zeta_1, \ldots, \zeta_n) = \Phi_\calW(\zeta_1, \ldots, \zeta_n, q^\omega \zeta_n, \ldots, q^\omega \zeta_1),
\end{equation*}
where we treat $\widetilde \Psi_n$ as a mapping from $\bbC^n$ to $\calW = V^{\otimes n} \otimes V^{\otimes n}$, see figure \ref{f:asl2}. Now the graphical equation \ref{f:fesl2d} can be represented as the equation given in figure \ref{f:rrqkzsl2}.
We come to the following analytical equation
\begin{multline*}
\Phi_\calW(\zeta_1, \ldots, q^\omega \zeta_n, q^{2 \omega} \zeta_n, \ldots, q^\omega \zeta_1) \\*
= \check R_{V | V}^{(n + 1, \, n + 2)}(q^\omega \zeta_{n - 1} | q^{2 \omega} \zeta_n) \ldots \check R_{V | V}^{(2 n - 1, \, 2 n)}(q^\omega \zeta_1 | q^{2 \omega} \zeta_n) P_\lambda \Delta^{(1)}(\zeta_n) \\*
\check R_{V | V}^{(1, \, 2)}(\zeta_1 | \zeta_n) \ldots \check R_{V | V}^{(n - 1, \, n)}(\zeta_{n - 1} | \zeta_n) 
\Phi_\calW(\zeta_1, \ldots, \zeta_n, q^\omega \zeta_n, \ldots, q^\omega \zeta_1),
\end{multline*}
where
\begin{equation}
\Delta(\zeta) = d_{V | V}^{n - 1} \, \phi^\alpha_V(\zeta) (\widetilde X^{-1})^t \widetilde X A^\alpha_V, \label{dz}
\end{equation}
and $P_\lambda$ is the left shift of the tensor product, see the appendix.  Using the representation~(\ref{pl}) and equations (\ref{pdpr}), one can transform this equation to the form
\begin{multline}
\Phi_\calW(\zeta_1, \ldots, q^\omega \zeta_n, q^{2 \omega} \zeta_n, \ldots, q^\omega \zeta_1) \\*
= R^{(n + 2, \, n + 1)}_{V | V}(q^\omega \zeta_{n - 1} | q^{2 \omega} \zeta_n) \ldots R^{(2 n, \, n + 1)}_{V | V}(q^\omega \zeta_1 | q^{2 \omega} \zeta_n) P^{(n, \, n + 1)} \Delta^{(n)} (\zeta_n) \\*
R^{(1, \, n)}_{V | V}(\zeta_1 | \zeta_n) \ldots R^{(n - 1, \, n)}_{V | V}(\zeta_{n - 1} | \zeta_n) \, \Phi_\calW(\zeta_1, \ldots, \zeta_n, q^\omega \zeta_n, \ldots, q^\omega \zeta_1). \label{rqkzb}
\end{multline}

Note that in the case under consideration $s \calW = \calW$  for any $s \in \rmS_N$ and the mappings $\Phi_{s \calW}$ differ only by a permutation of arguments. There is only one independent mapping $\Delta_i(\zeta)$ and only one independent equation~(\ref{qkzb}). It is evident that equation (\ref{rqkzb}) is equivalent to the equation
\begin{multline}
\check R^{(n, \, n + 1)}_{V | V} (q^\omega \zeta_n | q^{2 \omega} \zeta_n) \Phi_\calW(\zeta_1, \ldots, q^\omega \zeta_n, q^{2 \omega} \zeta_n, \ldots, q^\omega \zeta_1) \\*
=  \Lambda_{\calW n}(\zeta_1, \ldots, \zeta_n, q^\omega \zeta_n, \ldots, q^\omega \zeta_1) \Phi_\calW(\zeta_1, \ldots, \zeta_n, q^\omega \zeta_n, \ldots, q^\omega \zeta_1), \label{lprp}
\end{multline}
where the mapping $\Lambda_{\calW n}$ is defined by equation (\ref{lwia}) with $p = q^{2 \omega}$ and the mapping $\Delta_n$ coinciding with the mapping $\Delta$ given by equation (\ref{dz}). Assume that $\Phi_\calW$ satisfies the qKZ equation (\ref{qkzb}) for $i = n$ and with the same $p$ and $\Delta_n$. In this case (\ref{lprp}) turns exactly into this qKZ equation. Note that we include equalities (\ref{rd}) into the definition of the qKZ equation. It worth to remark that we have only one relation of type (\ref{ddr}). Equation (\ref{aarraa}) and the invariance relation given in figure \ref{f:pi} guaranty its validity.

Summing up, we conclude that if the mapping $\Phi_\calW$ satisfies the qKZ equation
\begin{multline*}
\Phi_\calW(\eta_1, \ldots, q^{2 \omega} \eta_n, \ldots, \eta_{2 n}) \\= \check R^{(n, \, n + 1)}_{V | V}(\eta_{n + 1} | q^{2 \omega} \eta_n) \ldots \check R^{(2 n - 1, \, 2 n)}_{V | V}(\zeta_{2 n} | q^{2 \omega} \eta_n) P_\lambda \, \Delta^{(1)}(\eta_n) \\*
\check R^{(1, \, 2)}_{V | V}(\eta_1 | \eta_n) \ldots \check R^{(n - 1, \, n)}_{V | V}(\eta_{n - 1} | \eta_n) \Phi_\calW(\eta_1, \ldots, \eta_n, \ldots, \eta_{ 2 n})
\end{multline*}
with $\Delta$ defined by equation (\ref{dz}), then the mapping $\Psi_n(\zeta_1, \ldots, \zeta_n)$ determined by the equation
\begin{equation*}
\Psi_n(\zeta_1, \ldots, \zeta_n)^{i_1 \ldots i_n}{}_{j_1 \ldots j_n} = \sum_{k_1, \ldots, k_n} \Phi_\calW(\zeta_1, \ldots, \zeta_n, q^\omega \zeta_n, \ldots, q^\omega \zeta_1)^{i_1 \ldots i_n k_n \ldots k_1} \widetilde X_{k_n j_n} \ldots \widetilde X_{k_1 j_1}
\end{equation*}
satisfies the reduced qKZ equation depicted in figure \ref{f:fesl2}. Using the invariance relation given in figure \ref{f:pi}, one can easily demonstrate that equations (\ref{rpr}) are also satisfied. 

As an example we consider the quantum loop algebra $\uqlslii$ and its $m$-dimensional irreducible representation $\varphi^m = \varphi^{(m / 2, \, - m / 2)}$ generated by the Jimbo's homomorphism from the representation $\pi^{(m / 2, \, - m / 2)}$ of $\uqglii$, see the paper \cite{NirRaz19}. For this representation we have
\begin{align*}
& \varphi^m_\zeta(q^{\nu h_0}) = \sum_{i = 1}^{m + 1} q^{- \nu(m - 2 i + 2)} E_{i i}, && \varphi^m_\zeta(q^{\nu h_1}) = \sum_{i = 1}^{m + 1} q^{\nu(m - 2 i + 2)} E_{i i}, \\
& \varphi^m_\zeta(e_0) = \zeta^{s_0} \sum_{i = 1}^m E_{i + 1, \, i}, && \varphi^m_\zeta(e_1) = \zeta^{s_1} \sum_{i = 1}^m \, [i]_q [m - i + 1]_q E_{i, \, i + 1}, \\
& \varphi^m_\zeta(f_0) = \zeta^{- s_0} \sum_{i = 1}^m \, [i]_q [m - i + 1]_q E_{i, \, i + 1}, && \varphi^m_\zeta(f_1) = \zeta^{- s_1} \sum_{i = 1}^m E_{i + 1, \, i}.
\end{align*}
Here and below we denote by $E_{i j}$, $i, j = 1, \ldots, m + 1$, the usual unit matrices. Using the definition of the dual representation (\ref{fsa}) and the equations
\begin{equation*}
S(q^x) = q^{- x}, \qquad S(e_i) = - q^{-h_i} e_i, \qquad S(f_i) = - f_i \, q^{h_i},
\end{equation*}
we obtain
\begin{align*}
& \varphi^{m*}_\zeta(q^{\nu h_0}) = \sum_{i = 1}^{m + 1} q^{\nu(m - 2 i + 2)} E_{i i}, \quad \varphi^{m*}_\zeta(q^{\nu h_1}) = \sum_{i = 1}^{m + 1} q^{- \nu(m - 2 i + 2)} E_{i i}, \\
& \varphi^{m*}_\zeta(e_0) = - \zeta^{s_0} \sum_{i = 1}^m q^{m - 2 i} E_{i, \, i + 1}, \quad \varphi^{m*}_\zeta(e_1) = - \zeta^{s_1} \sum_{i = 1}^m q^{-(m - 2 i + 2)} [i]_q [m - i + 1]_q E_{i + 1, \, i}, \\
& \varphi^{m*}_\zeta(f_0) = - \zeta^{- s_0} \sum_{i = 1}^m q^{-(m - 2 i)} [i]_q [m - i + 1]_q E_{i + 1, \, i}, \quad \varphi^{m*}_\zeta(f_1) = - \zeta^{- s_1} \sum_{i = 1}^m q^{m - 2 i + 2} E_{i, \, i + 1}.
\end{align*}
One can demonstrate that
\begin{equation*}
\varphi^{m *}_\zeta(a) = O\strut_{V^m} \, \varphi_{q^{- 2/s} \zeta}(a) \, O_{V^m}^{-1},
\end{equation*}
where
\begin{equation*}
O_{V^m} = \sum_{i = 1}^{m+1} (-1)^{m - i + 1} q^{(m - i + 1)(2 - 2 s_0 / s -i)} E_{m - i + 2, \, i}.
\end{equation*}
Thus, in the case of the representation $\varphi^m_\zeta$ we have $\delta = - 2/s$. Since in this case $\varepsilon = 4/s$, see (\ref{etts}), we have $\omega = 2/s$.

One can demonstrate that
\begin{equation*}
\rho^0_{V^m | V^m}(\zeta_1 | \zeta_2) = q^{ - m^2 / 2} \frac{(q^2 \zeta_{1 2}^s; \, q^4)^2_\infty}{(q^{2 m + 2} \zeta_{1 2}^s; \, q^4)_\infty \, (q^{- 2 m + 2} \zeta_{1 2}^s; \, q^4)_\infty},
\end{equation*}
see the paper \cite{IdzTokIohJimMiw93}. It follows that
\begin{equation*}
\rho^0_{V^m | V^m} (q^{-2/s} \zeta_1 | \zeta_2)^{-1} \rho^0_{V^m | V^m}(\zeta_1 | \zeta_2)^{-1} = q^{m^2} \prod_{i = 0}^{m - 1} \frac{1 - q^{- 2 m + 2 i} \zeta_{1 2}^s}{1 - q^{2 m - 2 i - 2} \zeta_{1 2}^s}.
\end{equation*}
Inserting 
\begin{equation*}
\kappa_{V^m | V^m}(\zeta_1 | \zeta_2) =(\zeta_{1 2}^s)^{m / 2} \frac{(q^{2 m + 2} \zeta_{1 2}^s; \, q^4)_\infty}{(q^{2 m + 2} \zeta_{1 2}^{-s}; \, q^4)_\infty} \frac{(q^2 \zeta_{1 2}^{-s}; \, q^4)_\infty}{(q^2 \zeta_{1 2}^s; \, q^4)_\infty}
\end{equation*}
into equation (\ref{rrkk}), we see that this equation is satisfied if
\begin{equation*}
d_{V^m | V^m} = (-1)^m.
\end{equation*}
Notice that in the case of an even $m = 2 k$ we have a rational expression
\begin{equation*}
\kappa_{V^{2 k} | V^{2 k}}(\zeta_1 | \zeta_2) =(\zeta_{1 2}^s)^k \prod_{i = 1}^k \frac{1 - q^{4 i - 2} \zeta_{1 2}^s}{1 - q^{4 i - 2} \zeta_{1 2}^{- s}}.
\end{equation*}

The $R$-operators with the obtained normalization appear in commutation relations of vertex operators \cite{IdzTokIohJimMiw93}.

\subsection{General case}

In the general case we again use the normalization (\ref{rkr}), where $V_1 = V_2 = V$ and the normalization factor $\rho^0_{V | V}(\zeta_1 | \zeta_2)$ is determined by the condition (\ref{rvv}). As the function $\kappa_{V | V}(\zeta_1 | \zeta_2)$ we take a solution to the difference equation
\begin{equation}
\rho^0_{V | V}(q^{-\varepsilon} \zeta_1 | \zeta_2)^{-1} \rho^0_{V^{} | V_{}}(\zeta_1 | \zeta_2) \, \kappa_{V^{} | V^{}}(q^{-\varepsilon} \zeta_1 | \zeta_2)^{-1} \kappa_{V^{} | V^{}}(\zeta_1 | \zeta_2) = d_{V | V} \label{rrkkgc}
\end{equation}
for some complex constant $d_{V | V}$. In this case the coefficients $D_{V |V}(\zeta_1 | \zeta_2)$ entering the crossing relations represented by figures \ref{f:cd} and \ref{f:ce} becomes equal to $d^{}_{V | V}$ and $d_{V | V}^{-1}$ respectively. Certainly, we again assume that the conditions (\ref{kvvkvv}) and (\ref{kvvo}) are satisfied.

\begin{figure}[t!]
\centering
\includegraphics{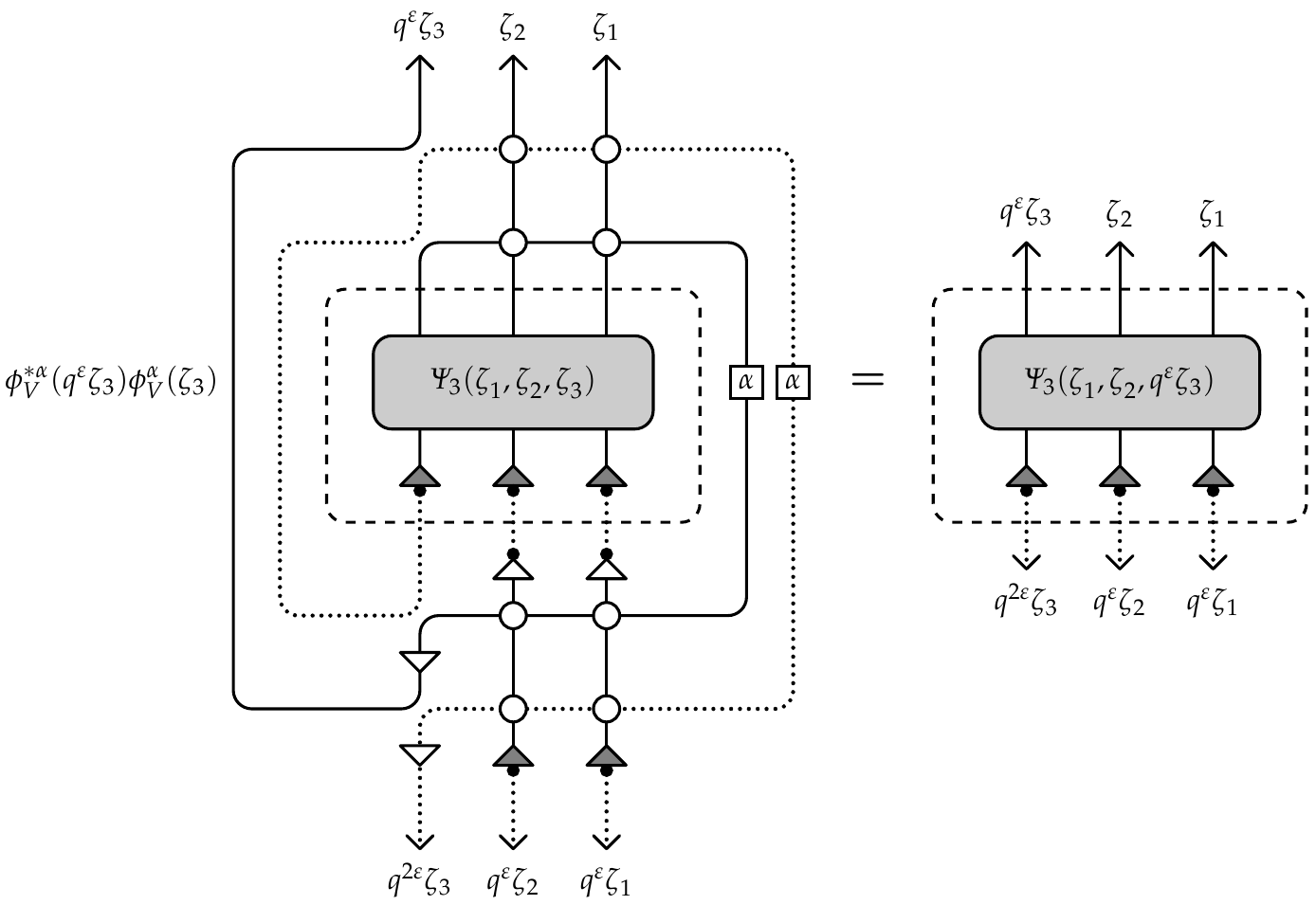}
\caption{}
\label{f:gca}
\end{figure}

\begin{figure}[t!]
\centering
\includegraphics{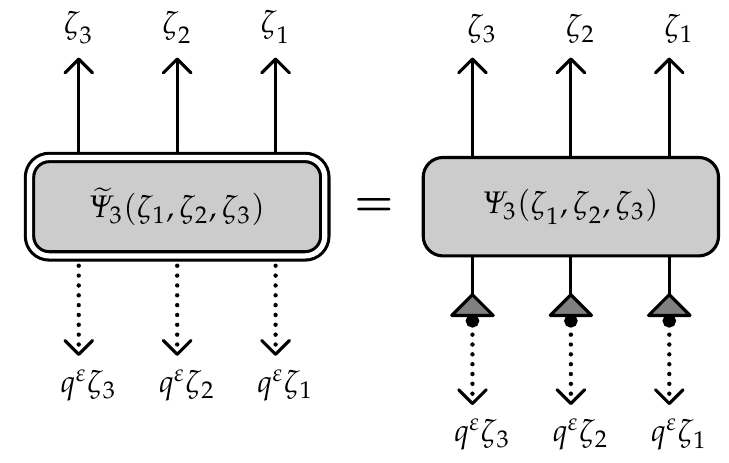}
\caption{}
\label{f:td}
\end{figure}

\begin{figure}[t!]
\centering
\includegraphics{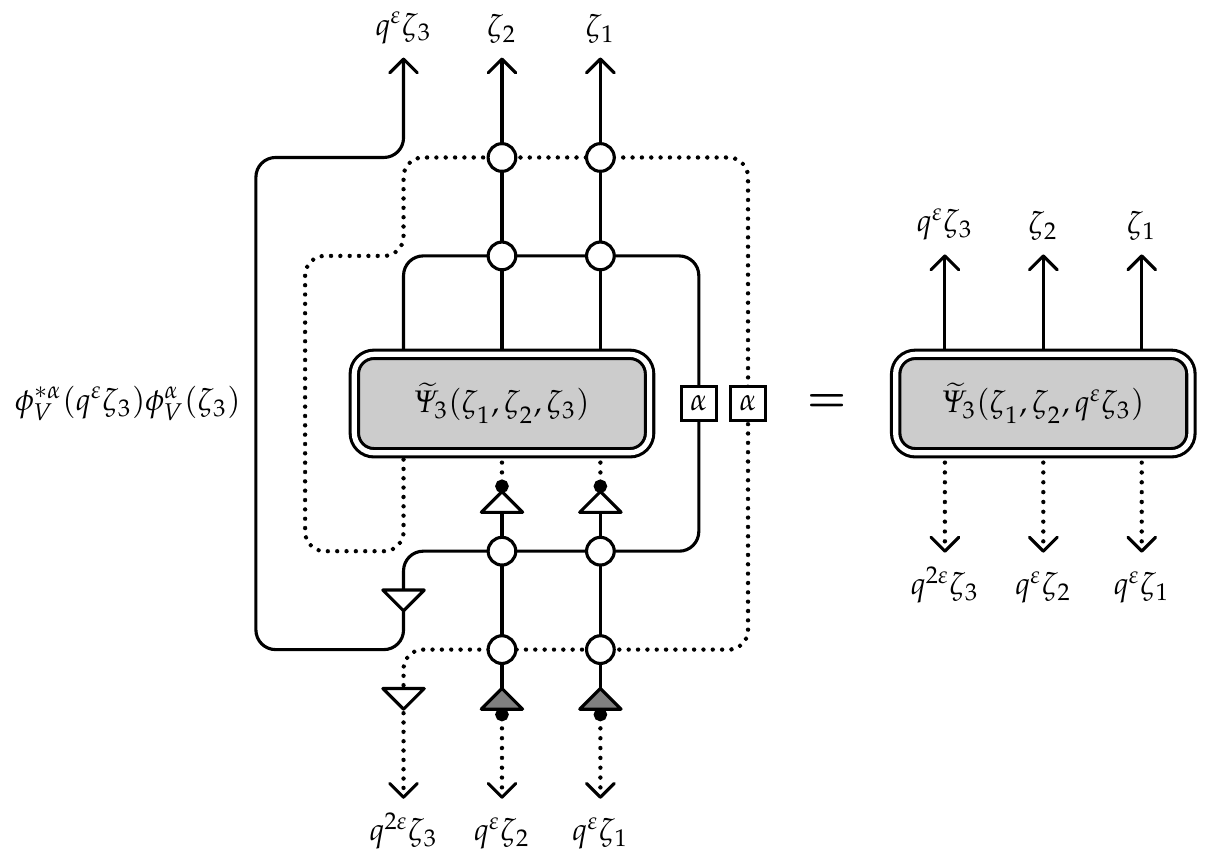}
\caption{}
\label{f:gcb}
\end{figure}

\begin{figure}[t!]
\centering
\includegraphics{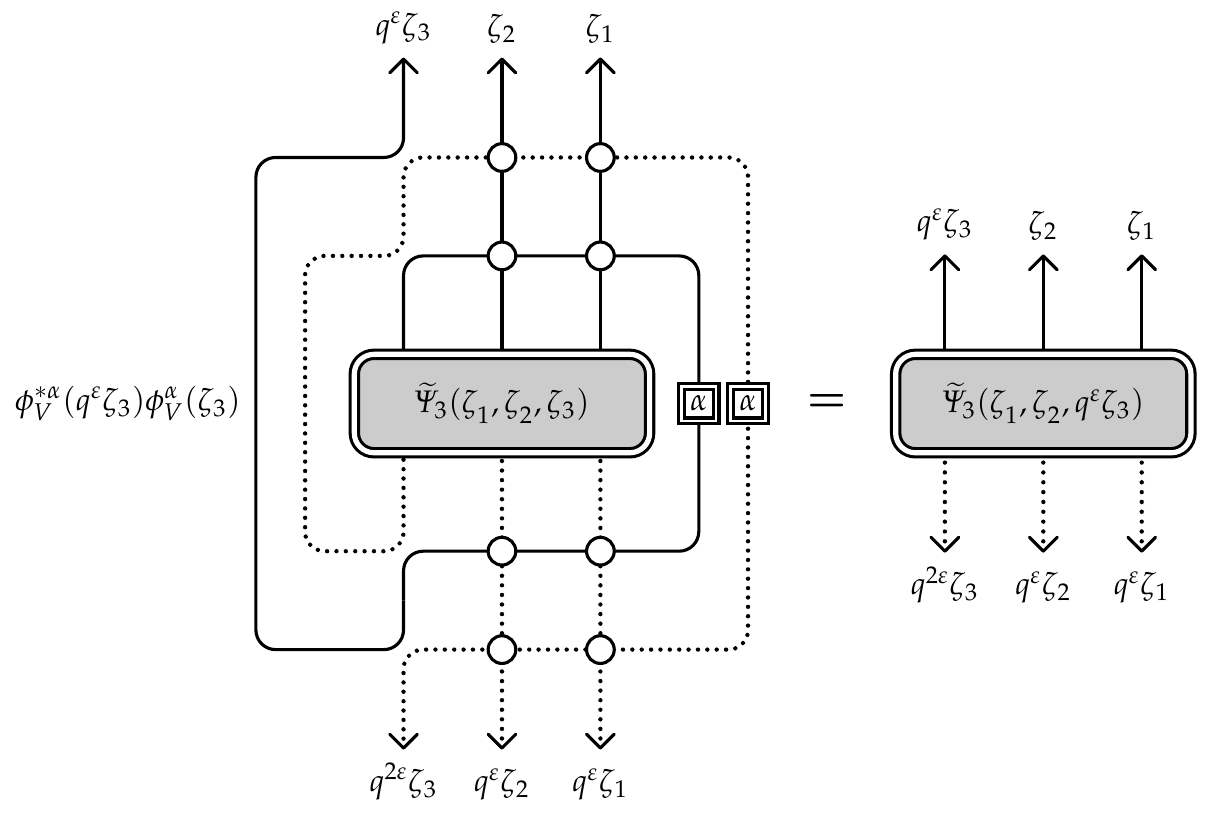}
\caption{}
\label{f:gcc}
\end{figure}

\begin{figure}[t!]
\centering
\begin{minipage}{0.45\textwidth}
\centering
\includegraphics{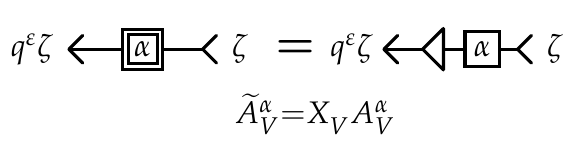}
\caption{}
\label{f:tga}
\end{minipage} \hfil
\begin{minipage}{0.45\textwidth}
\centering
\includegraphics{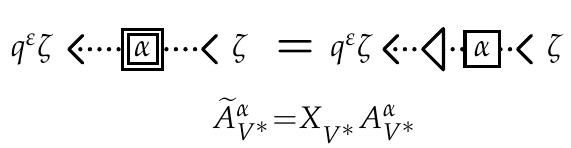}
\caption{}
\label{f:tgb}
\end{minipage}
\end{figure}

Now, we transform the graphical equation in figure \ref{f:fine} to the form which can be seen in figure \ref{f:gca}. It is worth noting that here we use the relation described by figure \ref{f:dox}. It is natural to introduce the object $\widetilde \Psi_n(\zeta_1, \ldots, \zeta_n)$ now defined by figure \ref{f:td}. The equation in figure \ref{f:gca} takes the form presented by figure \ref{f:gcb}. Using the invariance relation given in figure \ref{f:xi}, we move the leftmost upper white triangle to the right and then the leftmost lower white triangle also to the right. The resulting equation is represented by figure \ref{f:gcc}, where the modified twisting operators are described by figures \ref{f:tga} and \ref{f:tgb}.
\begin{figure}[t!]
\centering
\includegraphics{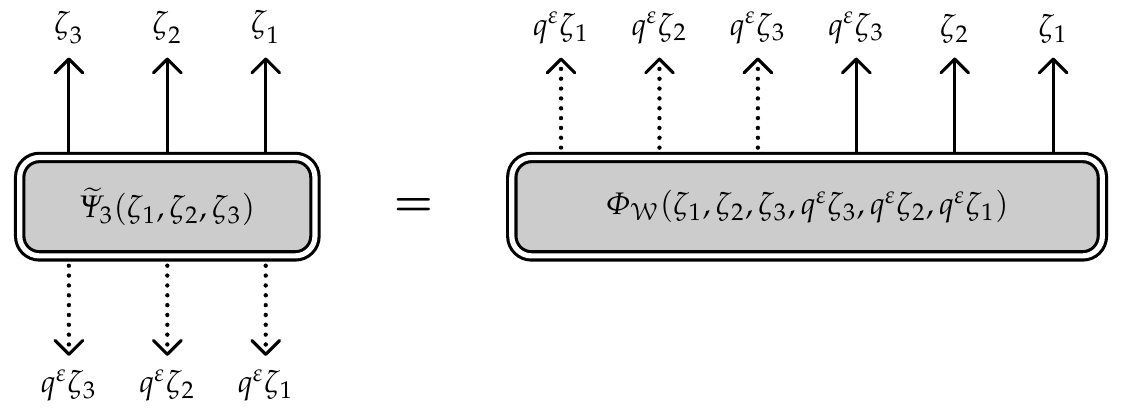}
\caption{}
\label{f:a}
\end{figure}

\begin{figure}[t!]
\centering
\includegraphics{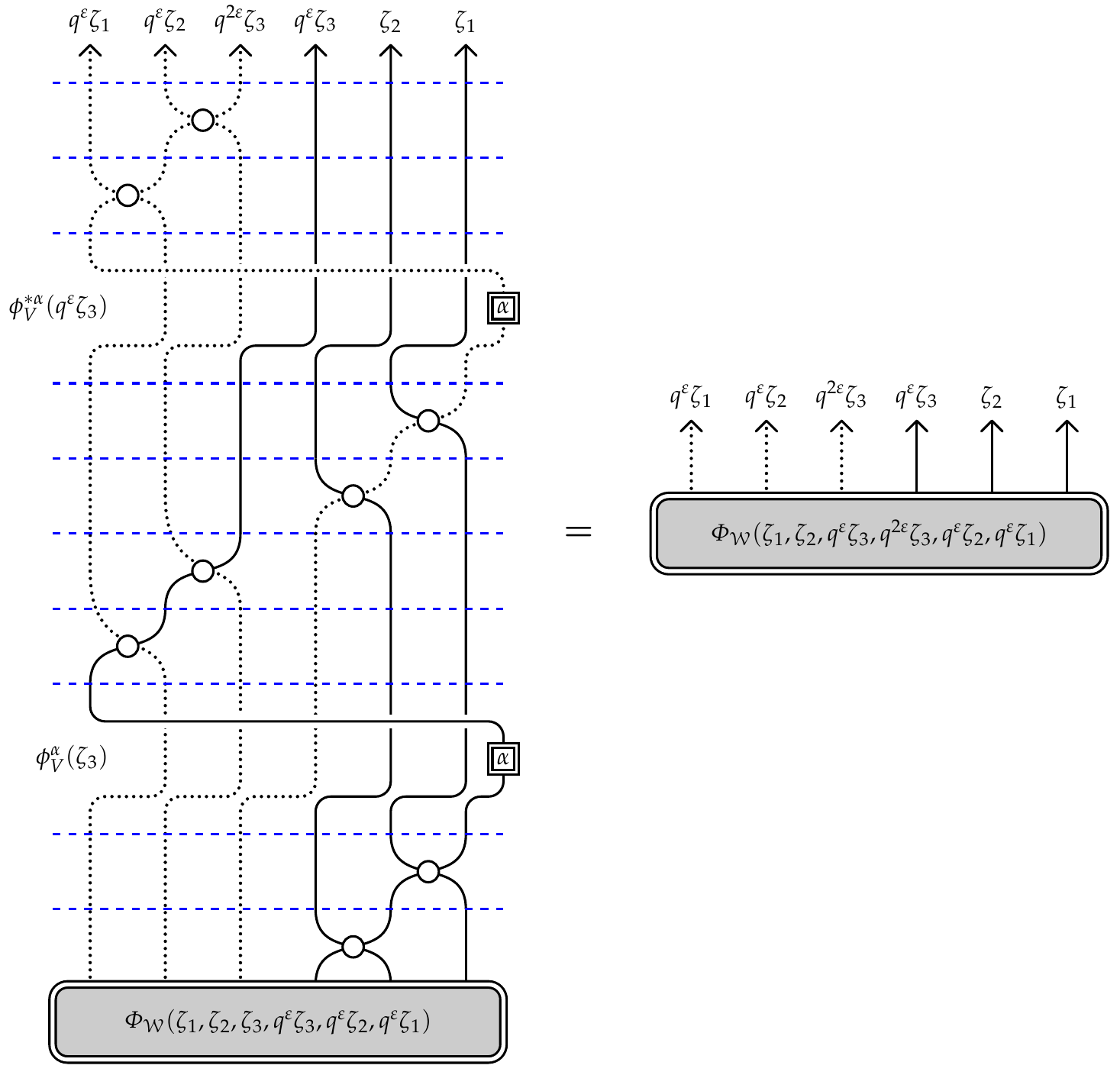}
\caption{}
\label{f:rqkz}
\end{figure}

As in the case of an almost self-dual representation considered in the previous section, one can generate solutions of the reduced qKZ equation from the solutions of the corresponding usual qKZ equation. Let $N = 2n$ and $\calW$ be the tensor product (\ref{www}), where $W_i = V$ for $i \in \interval{1}{n}$, $W_i = V^*$ for $i \in \interval{n + 1}{2 n}$, and $\Phi_\calW$ a mapping from $\bbC^{2 n}$ to $\calW$. Consider the ansatz
\begin{equation*}
\widetilde \Psi_n(\zeta_1, \ldots, \zeta_n) = \Phi_\calW(\zeta_1, \ldots, \zeta_n, q^\varepsilon \zeta_n, \ldots, q^\varepsilon \zeta_1)
\end{equation*}
where $\widetilde \Psi_n$ is considered as a mapping from $\bbC^n$ to $\calW = V^{\otimes n} \otimes V^{*\otimes n}$, see figure \ref{f:a}. Now the graphical equation given in figure \ref{f:gcc} can be represented as the graphical equation depicted in figure \ref{f:rqkz}. The analytical form of this equation is 
\begin{align}
\Phi_\calW(& \zeta_1, \ldots, \zeta_{n - 1}, q^\varepsilon \zeta_n, q^{2 \varepsilon} \zeta_n, q^\varepsilon \zeta_{n - 1}, \ldots, q^\varepsilon \zeta_n) \notag \\*
= {} & \check R^{(n + 1, \, n + 2)}_{V^* | V^*}(q^\varepsilon \zeta_{n - 1} | q^{2 \varepsilon} \zeta_n) \ldots \check R^{(2 n - 1, \, 2 n)}_{V^* | V^*}(q^\varepsilon \zeta_1 | q^{2 \varepsilon} \zeta_n) P_\lambda \Delta^{*(1)}(q^\varepsilon \zeta_n) \notag \\*
& \hspace{1em} \check R^{(1, \, 2)}_{V | V^*}(\zeta_1 | q^\varepsilon \zeta_n) \ldots \check R^{(n - 1, \, n)}_{V | V^*}(\zeta_{n - 1} | q^\varepsilon \zeta_n) \notag \\*
& \check R^{(n + 1, \, n + 2)}_{V^* | V}(q^\varepsilon \zeta_{n - 1} | q^\varepsilon \zeta_n) \ldots \check R^{(2 n - 1, \, 2 n)}_{V^* | V}(q^\varepsilon \zeta_1 | q^\varepsilon \zeta_n) P_\lambda \Delta^{(1)}(\zeta_n) \notag \\*
& \hspace{1em}  \check R^{(1, \, 2)}_{V | V}(\zeta_1 | \zeta_n) \ldots \check R^{(n - 1, \, n)}_{V | V}(\zeta_{n - 1} | \zeta_n)  \Phi_\calW(\zeta_1, \ldots, \zeta_{n - 1}, \zeta_n, q^\varepsilon \zeta_n, q^\varepsilon \zeta_{n - 1}, \ldots, q^\varepsilon \zeta_1), \label{pwcr}
\end{align}
where
\begin{equation}
\Delta(\zeta) = \phi\strut^\alpha_V(\zeta) X\strut_V A\strut^\alpha_V, \qquad \Delta^*(\zeta) = \phi\strut^{* \alpha}_V(\zeta) X\strut_{V^*} A\strut^\alpha_{V^*}. \label{dds}
\end{equation}
 Using the representation~(\ref{pl}) and equations (\ref{pdpr}), one can transform this equation to the form
\begin{align}
\Phi_\calW(& \zeta_1, \ldots, \zeta_{n - 1}, q^\varepsilon \zeta_n, q^{2 \varepsilon} \zeta_n, q^\varepsilon \zeta_{n - 1}, \ldots, q^\varepsilon \zeta_n) \notag \\*
= {} & R^{(n + 2, \, n + 1)}_{V^* | V^*}(q^\varepsilon \zeta_{n - 1} | q^{2 \varepsilon} \zeta_n) \ldots R^{(2 n, \, n + 1)}_{V^* | V^*}(q^\varepsilon \zeta_1 | q^{2 \varepsilon} \zeta_n) P^{(n, \, n + 1)} \Delta^{*(n)}(q^\varepsilon \zeta_n) \notag \\*
& \hspace{1em} R^{(1, \, n)}_{V | V^*}(\zeta_1 | q^\varepsilon \zeta_n) \ldots R^{(n - 1, \, n)}_{V | V^*}(\zeta_{n - 1} | q^\varepsilon \zeta_n) \notag \\
& R^{(n + 2, \, n + 1)}_{V^* | V}(q^\varepsilon \zeta_{n - 1} | q^\varepsilon \zeta_n) \ldots R^{(2 n, \, n + 1)}_{V^* | V}(q^\varepsilon \zeta_1 | q^\varepsilon \zeta_n) P^{(n, \, n + 1)} \Delta^{(n)}(\zeta_n) \notag  \\*
& \hspace{1em} R^{(1, \,  n)}_{V | V}(\zeta_1 | \zeta_n) \ldots R^{(n - 1, \, n)}_{V | V}(\zeta_{n - 1} | \zeta_n) \Phi_\calW(\zeta_1, \ldots, \zeta_{n - 1}, \zeta_n, q^\varepsilon \zeta_n, q^\varepsilon \zeta_{n - 1}, \ldots, q^\varepsilon \zeta_1). \label{pwr}
\end{align}

Inserting in the middle of the right hand side of equation (\ref{pwcr}) the identity
\begin{equation*}
\check R_{V | V^*}^{(n, \, n + 1)}(q^\varepsilon \zeta_n | q^\varepsilon \zeta_n) \check R_{V^* | V}^{(n, \, n + 1)}(q^\varepsilon \zeta_n | q^\varepsilon \zeta_n) = \id^{(n, \, n + 1)}_{V^* | V},
\end{equation*}
we come to the equation
\begin{multline}
\Phi_\calW(\zeta_1, \ldots, q^\varepsilon \zeta_n, q^{2 \varepsilon} \zeta_n, \ldots, q^\varepsilon \zeta_1) \\
\Lambda_{\calW, n + 1}(\zeta_1, \ldots, q^\varepsilon \zeta_n, q^\varepsilon \zeta_n, \ldots, q^\varepsilon \zeta_1) \Lambda_{\calW n}(\zeta_1, \ldots, \zeta_n, q^\varepsilon \zeta_n, \ldots, q^\varepsilon \zeta_1) \\ \Phi_\calW(\zeta_1, \ldots, \zeta_n, q^\varepsilon \zeta_n, \ldots, q^\varepsilon \zeta_1), \label{pll}
\end{multline}
where the mappings $\Lambda_{\calW n}$ and $\Lambda_{\calW, n + 1}$ are defined by equation (\ref{lwia}) with $p = q^\varepsilon$ and the mappings $\Delta_n$ and $\Delta_{n + 1}$ coincide with the mappings $\Delta$ and $\Delta^*$ determined by equation (\ref{dds}). Note that in the case under consideration, there are only two independent mappings $\Delta_i$ related to $V$ and $V^*$. Respectively, there are also only two independent equations (\ref{qkzb}). It is clear that if $\Phi_\calW$ satisfies the independent qKZ equations (\ref{qkzb}) for $i = n$ and $i = n + 1$, it satisfies equation (\ref{pll}), and, that is equivalent, equation (\ref{pwcr}) or equation (\ref{pwr}). Now we have four independent relations of type (\ref{ddr}). Their validity follows from equation (\ref{aarraa}) and the invariance relation given in figure \ref{f:xi}.

Summing up, we conclude that if the mapping $\Phi_\calW$ satisfies the qKZ equations
\begin{multline*}
\Phi_\calW(\eta_1, \ldots, q^\varepsilon \eta_n, \ldots, \eta_{2 n}) \\ = \check R^{(n, \, n + 1)}_{V^* | V}(\eta_{n + 1} | q^\varepsilon \eta_n) \ldots \check R^{(2 n - 1, \, 2 n)}_{V^* | V}(\zeta_{2 n} | q^\varepsilon \eta_n) P_\lambda \, \Delta^{(1)}(\eta_n) \\*
\check R^{(1, \, 2)}_{V | V}(\eta_1 | \eta_n) \ldots \check R^{(n - 1, \, n)}_{V | V}(\eta_{n - 1} | \eta_n) \Phi_\calW(\eta_1, \ldots, \eta_n, \ldots, \eta_{ 2 n})
\end{multline*}
and
\begin{multline*}
\Phi_\calW(\eta_1, \ldots, q^\varepsilon \eta_{n + 1}, \ldots, \eta_{2 n}) \\ = \check R^{(n - 1, \, n)}_{V^* | V^*}(\eta_{n + 2} | q^\varepsilon \eta_{n + 1}) \ldots \check R^{(2 n - 1, \, 2 n)}_{V^* | V^*}(\zeta_{2 n} | q^\varepsilon \eta_{n + 1}) P_\lambda \, \Delta^{(1)}(\eta_n) \\*
\check R^{(1, \, 2)}_{V | V^*}(\eta_1 | \eta_{n + 1}) \ldots \check R^{(n, \, n + 1)}_{V | V^*}(\eta_n | \eta_{n + 1}) \Phi_\calW(\eta_1, \ldots, \eta_n, \ldots, \eta_{ 2 n})
\end{multline*}
with $\Delta$ and $\Delta^*$ defined by equation (\ref{dds}), then the mapping $\Psi_n(\zeta_1, \ldots, \zeta_n)$ determined by the equation
\begin{equation*}
\Psi_n(\zeta_1, \ldots, \zeta_n)^{i_1 \ldots i_n}{}_{j_1 \ldots j_n} = \sum_{k_1, \ldots, k_n} \Phi_\calW(\zeta_1, \ldots, \zeta_n, q^\varepsilon \zeta_n, \ldots, q^\varepsilon \zeta_1)^{i_1 \ldots i_n}{}_{k_n \ldots k_1} X^{k_n}{}_{j_n} \ldots X^{k_1}{}_{j_1}
\end{equation*}
satisfies the reduced qKZ equation depicted in figure \ref{f:fine}. Using the invariance relation given in figure \ref{f:xi}, one can easily demonstrate that equations (\ref{rpr}) are also satisfied.

As an example consider the case of $\uqlsllpo$. In this case $\varepsilon = 2(l + 1)/s$. Assume that $V = \bbC^{l + 1}$ and $\varphi$ is the representation of $\uqlsllpo$ generated from the first fundamental representation of $\uqgllpo$ by the Jimbo's homomorphism \cite{Jim86a}. The normalization factor $\rho^0_{V | V}(\zeta_1 | \zeta_2)$ defined by the condition (\ref{rvv}) has the form
\begin{equation*}
\rho^0_{V | V}(\zeta_1 | \zeta_2) = q^{-l/l+1} \exp(F_{l + 1}(q^{-l} \zeta_{1 2}^s) - F_{l + 1}(q^l \zeta_{1 2}^s)),
\end{equation*}
where the function $F_m(\zeta)$ is given by the equation
\begin{equation*}
F_m (\zeta) = \sum_{n = 1}^\infty \frac{1}{[m]_{q^n}} \frac{\zeta^n}{n},
\end{equation*}
see the paper \cite{NirRaz19}. One can demonstrate that
\begin{equation*}
\rho^0_{V | V}(\zeta_1 | \zeta_2) = q^{-l/(l + 1)} \frac{(q^2 \zeta_{12}^s; \, q^{2(l + 1)})_\infty}{(\zeta_{12}^s; \, q^{2(l + 1)})_\infty} \frac{(q^{2 l} \zeta_{12}^s; \, q^{2(l + 1)})_\infty}{(q^{2(l + 1)} \zeta_{12}^s; \, q^{2(l + 1)})_\infty},
\end{equation*}
and obtain
\begin{equation*}
\rho^0_{V | V}(q^{- 2(l + 1)/s} \zeta_1 | \zeta_2)^{-1} \rho^0_{V^{} | V_{}}(\zeta_1 | \zeta_2) = \frac{1 - q^{-2} \zeta_{1 2}^s}{1 - \zeta_{1 2}^s} \frac{1 - q^{-2 l} \zeta_{1 2}^s}{1 - q^{-2(l + 1)} \zeta_{1 2}^s}.
\end{equation*}
Inserting
\begin{equation*}
\kappa_{V | V}(\zeta_1 | \zeta_2) = (\zeta_{1 2}^s)^{l/(l + 1)} \frac{(q^2 \, \zeta_{1 2}^{-s}; \, q^{2(l + 1)})_\infty}{(q^2 \, \zeta_{1 2}^s; \, q^{2(l + 1)})_\infty}  \frac{(q^{2(l + 1)} \zeta_{1 2}^s; \, q^{2(l + 1)})_\infty}{(q^{2( l + 1)} \zeta_{1 2}^{-s}; \, q^{2(l + 1)})_\infty}
\end{equation*}
into equation (\ref{rrkkgc}), we see that this equation is satisfied if $d_{V | V} = 1$. It is worth to note that the $R$-operators defined with such choice a of $\kappa_{V | V}(\zeta_1 | \zeta_2)$ enter commutation relations for quantum vertex operators, see, for example, the paper \cite{Koj18}, where the most general supersymmetric case is considered.

\section{Conclusions}

In the paper \cite{KluNirRaz19a}, in continuation of the previous works \cite{Auf11, AufKlu12, BooHutNir18,  RibKlu19}, a difference-type functional equation for the zero temperature inhomogeneous reduced density operator of a quantum integrable vertex model related to an arbitrary complex simple Lie algebra was derived. By analogy with the simplest case of the quantum loop algebra $\uqlslii$, this equation was called the reduced qKZ equation. However, for the general case, the real relationship with the qKZ equation remained unclear. This question was analyzed in the present paper.

There are two different cases. The first situation arises when the representation used to formulate the integrable system under consideration is almost self-dual. By this we mean that the dual representation is isomorphic to the initial representation up to a redefinition of the spectral parameter. In the second situation we deal with a general representation. It was demonstrated that in both cases a solution of the corresponding qKZ equation generate a solution to the reduced qKZ equation.

We believe that the results obtained in the present paper will serve as an additional incentive for the study of the qKZ equation and discovering new methods to find its solutions.

{\em Acknowledgments\/} This work was supported in part by the RFBR grant \#~20-51-12005. The author would like to thank the Max Planck Institute for Mathematics in Bonn, where this work was started, for the hospitality extended to him during his stay there in Febru\-ary-May 2019. 

\appendix

\setcounter{section}{1}
\setcounter{equation}{0}

\section*{Appendix. Quantum Knizhnik--Zamolodchikov equation} \label{s:qkze}

We define the symmetric group $\rmS_N$ on $N$ elements as the set of all mapping of the set $\{ 1, \, 2, \, \ldots, \, N \}$ onto itself. It is useful to interpret the elements of $\rmS_N$ as permutations in the following way. Let we have $N$ different objects occupying $N$ consecutive positions. We interpret $s \in \rmS_N$ as the permutation at which the object occupying the $i$th position goes to the $s(i)$th position. For example, the cyclic left shift of the objects is identified with $\lambda \in \rmS_N$ for which
\begin{equation*}
\lambda(1) = N, \qquad \lambda(i) = i - 1, \quad i = 2, \ldots, N.
\end{equation*}
The group $\rmS_N$ can be described as the group with generators $\sigma_i$, $i \in \interval{1}{N - 1}$, and relations
\begin{align}
& \sigma_i^2 = 1, && 1 \le i \le N - 1, \label{dra} \\
& (\sigma_i \sigma_j)^2 = 1, && 1 \le i < j \le N - 1, \quad j - i > 1, \label{drb} \\
& (\sigma_i \sigma_{i + 1})^3 = 1, && 1 \le i < N - 1, \label{drc}
\end{align}
see, for example, the book \cite{Hum96}. Here $\sigma_i$ is the permutation which swaps the objects at the $i$\,th and $(i + 1)$th positions.

Let $W_1$, \ldots, $W_N$ be finite dimensional vector spaces. Consider the tensor product
\begin{equation}
\calW = W_1 \otimes \cdots \otimes W_N. \label{www}
\end{equation}
Given $s \in \rmS_N$, apply the corresponding permutation to the factors of $\calW$ and denote the result as $s \calW$. In accordance with the identification of the elements of $\rmS_N$ with permutations described above, we have
\begin{equation*}
s \calW = (s \calW)_1 \otimes \cdots \otimes (s \calW)_N = W_{s^{-1}(1)} \otimes \cdots \otimes W_{s^{-1}(N)}.
\end{equation*}
For any $s, t \in S_n$ we define a linear mapping $P_s \colon t \calW \to s t \calW$ as follows
\begin{equation*}
P_s (v_1 \otimes \cdots \otimes v_N) = v_{s^{-1}(1)} \otimes \cdots \otimes v_{s^{-1}(N)},
\end{equation*}
where $v_i \in (t \calW)_i$ for all $i \in \interval{1}{N}$. It is clear that
\begin{equation*}
P_{s_1} P_{s_2} = P_{s_1 s_2}
\end{equation*}
for any $s_1, t_2 \in \rmS_N$. We use the notation
\begin{equation*}
P^{(i, \, i + 1)} = P_{\sigma_i}.
\end{equation*}

Let $W_1$, $\ldots$, $W_N$ be $\uqlg$-modules. We assume that for any distinct $i, j \in \interval{1}{N}$ the corresponding $R$-operators satisfy the unitarity relation of the form
\begin{equation}
\check R_{W_i | W_j}(\eta_i | \eta_j) \check R_{W_j | W_i}(\eta_j | \eta_i) = \id_{W_j \otimes W_i}, \label{aur} 
\end{equation}
where $\eta_i$ and $\eta_j$ are the spectral parameters associated with the modules $W_i$ and $W_j$ respectively. Certainly, the Yang--Baxter equation
\begin{equation}
R^{(1, \, 2)}_{W_i | W_j}(\eta_i | \eta_j) R^{(1, \, 3)}_{W_i | W_k}(\eta_i | \eta_k) R^{(2, \, 3)}_{W_j | W_k}(\eta_j | \eta_k) =  R^{(2, \, 3)}_{W_j | W_k}(\eta_j | \eta_k) R^{(1, \, 3)}_{W_i | W_k}(\eta_i | \eta_k) R^{(1, \, 2)}_{W_i | W_j}(\eta_i | \eta_j) \label{aybe}
\end{equation}
is also satisfied on $W_i \otimes W_j \otimes W_k$ for any distinct $i, j, k \in \interval{1}{N}$.

By construction, $\calW$ is an $\underbracket[.6pt]{\uqlg \otimes \cdots \otimes \uqlg}_N$-module. Let $\Phi_\calW$ be a mapping from $(\bbC^\times)^N$ to $\calW$. Given $s \in \rmS_N$, construct a mapping $\Phi_{s \calW}$ from $(\bbC^\times)^N$ to $s \calW$ in the following way. Represent $s$ as
\begin{equation}
s = \sigma_{i_m} \ldots \sigma_{i_1} \label{ssigma}
\end{equation}
and denote
\begin{equation*}
s_0 = e, \qquad s_k = \sigma_{i_k} \ldots \sigma_{i_1}, \quad 1 \le k \le m,
\end{equation*}
so that $s_m = s$. Assume that
\begin{equation}
\Phi_{s_0 \calW}(\eta_1, \ldots, \eta_N) = \Phi_\calW(\eta_1, \ldots, \eta_N)
\end{equation}
and define $\Phi_{s_k \calW} \colon \bbC^N \to s_k \calW$ for $k \in \interval{1}{m}$ by the recurrence relation
\begin{equation*}
\Phi_{s_k \calW} (\eta_{s_k^{-1}(1)}, \ldots, \eta_{s_k^{-1}(N)}) = \check R^{(i_k, \, i_k + 1)}_{(s_{k - 1} \calW)_{i_k} | (s_{k - 1} \calW)_{i_k + 1}}(\eta_{s_{k - 1}^{-1}(i_k)} | \eta_{s_{k - 1}^{-1}(i_{k + 1})}) \Phi_{s_{k - 1} \calW}(\eta_1, \ldots, \eta_N).
\end{equation*}
It follows from the defining relations (\ref{dra})--(\ref{drc}), the unitarity relations (\ref{aur}) and the Yang--Baxter equations (\ref{aybe}) that $\Phi_{s \calW} = \Phi_{s_m \calW}$ does not depend on the choice of the representation (\ref{ssigma}). It is clear that
\begin{multline}
\check R^{(i, \, i + 1)}_{(s \calW)_{s^{-1}(i)} | (s \calW)_{s^{-1}(i + 1)}}(\eta_{s^{-1}(i)} | \eta_{s^{-1}(i + 1)}) \Phi_{s \calW}(\eta_{s^{-1}(1)}, \ldots, \eta_{s^{-1}(i)}, \eta_{s^{-1}(i + 1)}, \ldots, \eta_{s^{-1}(N)}) \\= \Phi_{\sigma_i s \calW}(\eta_{s^{-1}(1)}, \ldots, \eta_{s^{-1}(i + 1)}, \eta_{s^{-1}(i)}, \ldots, \eta_{s^{-1}(N)}). \label{rd}
\end{multline}
for any $s \in \rmS_N$ and $i \in \interval{1}{N - 1}$.

For each $i = 1, \ldots N$ we choose a mapping $\Delta_i \colon \bbC \to \Aut(W_i)$ such that
\begin{equation}
(\Delta_j(\eta_j) \otimes \Delta_k(\eta_k)) R_{W_j | W_k}(\eta_j | \eta_k) = R_{W_j | W_k}(\eta_j | \eta_k) (\Delta_j(\eta_j) \otimes \Delta_k(\eta_k)) \label{ddr}
\end{equation}
for all distinct $j$ and $k$. This equation is used to prove the consistency of the quantum qKZ equations \cite{FreRes92}. In fact, the authors of the paper \cite{FreRes92} use the operators which do not depend on spectral parameters. However, the qKZ equations are consistent in our more general case as well.

Fix $i \in \interval{1}{N}$ and assume that
\begin{multline}
\Phi_{W_1 \otimes \cdots \otimes \widehat W_i \otimes \cdots \otimes W_N \otimes W_i}(\eta_1, \ldots, \widehat{\eta_i}, \ldots, \eta_N, p \eta_i) \\*
= P^{}_\lambda \Delta^{(1)}_i(\eta_i) \Phi_{W_i \otimes W_1 \otimes \cdots \otimes \widehat{W_i} \otimes \cdots \otimes W_N}(\eta_i, \eta_1, \ldots, \widehat{\eta_i}, \ldots, \eta_N) \label{pdp}
\end{multline}
for some complex parameter $p$. It follows from this equation that
\begin{multline}
\Phi_{(s \calW)_1 \otimes \cdots \otimes \widehat{(s \calW)_i} \otimes \cdots \otimes(s \calW)_N \otimes (s \calW)_i}(\eta_{s^{-1}(1)}, \ldots, \widehat{\eta_{s^{-1}(i)}}, \ldots, \eta_{s^{-1}(N)}, p \eta^{}_i) \\
= P^{}_\lambda \Delta^{(1)}_i(\eta_i) \Phi_{(s \calW)_i \otimes (s \calW)_1 \otimes \cdots \otimes \widehat{(s \calW)_i} \otimes \cdots \otimes (s \calW)_N}(\eta^{}_i, \eta_{s^{-1}(1)}, \ldots, \widehat{\eta_{s^{-1}(i)}}, \ldots, \eta_{s^{-1}(N)}) \label{pdps}
\end{multline}
for any $s \in \rmS_N$ such that $s(i) = i$.

Now, using equation (\ref{rd}), move the module $W_i$ in the tensor product (\ref{www}) to the first position. This gives the equation
\begin{multline*}
\Phi_{W_i \otimes W_1 \otimes \cdots \otimes \widehat{W_i} \otimes \cdots \otimes W_N}(\eta_i, \eta_1, \ldots \widehat{\eta_i}, \ldots, \eta_N) \\*
= \check R^{(1, \, 2)}_{W_1 | W_i}(\eta_1 | \eta_i) \ldots \check R^{(i - 1, \, i)}_{W_{i - 1} | W_i}(\eta_{i - 1} | \eta_i) \Phi_{W_1 \otimes \cdots \otimes W_i \otimes \cdots \otimes W_N}(\eta_1, \ldots, \eta_i, \ldots, \eta_N).
\end{multline*}
Further, use (\ref{pdp}) to jump to the last position and, using again (\ref{rd}), return $W_i$ to its initial position. This leads to the equation
\begin{multline*}
\Phi_{W_1 \otimes \cdots \otimes W_i \otimes \cdots \otimes W_N}(\eta_1, \ldots, p \eta_i, \ldots, \eta_N) = \\
\check R^{(i, \, i + 1)}_{W_{i + 1} | W_i}(\eta_{i + 1} | p \eta_i) \ldots \check R^{(N - 1, \, N)}_{W_N | W_i}(\eta_N | p \eta_i) \Phi_{W_1 \otimes \cdots \otimes \widehat{W_i} \otimes \cdots \otimes W_N \otimes W_i}(\eta_1, \ldots, \widehat{\eta_i}, \ldots, \eta_N, p \eta_i).
\end{multline*}
All this results in the equation
\begin{equation}
\Phi_\calW(\eta_1, \ldots, p \eta_i, \ldots, \eta_N) = \Lambda_{\calW i} (\eta_1, \ldots, \eta_N) \Phi_\calW(\eta_1, \ldots, \eta_i, \ldots, \eta_N). \label{qkzb}
\end{equation}
where the mapping $\Lambda_{\calW i} \colon (\bbC^\times)^N \to \Aut(\calW)$ is defined as 
\begin{multline}
\Lambda_{\calW i} (\eta_1, \ldots, \eta_N) = \check R^{(i, \, i + 1)}_{W_{i + 1} | W_i}(\eta_{i + 1} | p \eta_i) \ldots \check R^{(N - 1, \, N)}_{W_N | W_i}(\eta_N | p \eta_i) \\*
 P_\lambda \, \Delta^{(1)}_i(\eta_i) \check R^{(1, \, 2)}_{W_1 | W_i}(\eta_1 | \eta_i) \ldots \check R^{(i - 1, \, i)}_{W_{i - 1} | W_i}(\eta_{i - 1} | \eta_i). \label{lwia}
\end{multline}
Using the representation
\begin{equation}
P_\lambda = P_{\sigma_{N - 1}} \ldots P_{\sigma_i}P_{\sigma_{i - 1}} \ldots P_{\sigma_1} = P^{(N -1, \, N)} \ldots P^{(i, \, i + 1)} P^{(i - 1, \, i)} \ldots P^{(1, \, 2)}, \label{pl}
\end{equation}
and the equations
\begin{equation}
P\strut_s \, \Delta_i^{(j)}(\eta_i) = \Delta_i^{(s(j))}(\eta_i) P\strut_s, \qquad P\strut_s R^{(i, \, j)}_{W_i | W_j}(\eta_i | \eta_j) =  R^{(s(i), \, s(j))}_{W_i | W_j}(\eta_i | \eta_j) P{}_s, \label{pdpr}
\end{equation}
we rewrite the definition of $\Lambda_{\calW i}$ in the form
\begin{multline*}
\Lambda_{\calW i} (\eta_1, \ldots, \eta_N) = R^{(i + 1, \, i)}_{W_{i + 1} | W_i}(\eta_{i + 1} | p \eta_i) \ldots R^{(N, \, i)}_{W_N | W_i}(\eta_N | p \eta_i) \\
\Delta^{(i)}_i (\eta_i) R^{(1, \, i)}_{W_1 | W_i}(\eta_1 | \eta_i) \ldots R^{(i - 1, \, i)}_{W_{i - 1} | W_i}(\eta_{i - 1} | \eta_i).
\end{multline*}
It is evident that this equation implies equation (\ref{pdp}). When $i$ runs the whole interval~$\interval{1}{N}$ we obtain a system of difference equations whose consistency is guaranteed by the relations~(\ref{ddr}). This is the system that we call the qKZ equation. Together with (\ref{rd}) this is the system equivalent to the original system derived by Frenkel and Reshetikhin \cite{FreRes92}. It is useful to realize that one can also consider the equivalent system for the mappings $\Phi_{s \calW}$, consisting of equations (\ref{pdps}) and (\ref{rd}).

\providecommand{\href}[2]{#2}


\begin{thebibliography}{10}

\bibitem{KluNirRaz19a}

A.~Kl\"umper, Kh.~S. Nirov, and A.~V. Razumov, \emph{Reduced {qKZ} equation:
  general case}, \href{http://dx.doi.org/10.1088/1751-8121/ab3b9e}{J. Phys. A:
  Math. Gen.} \textbf{53} (2019), 015202 (35pp),
  \href{http://arxiv.org/abs/1905.06014}{{\tt arXiv:1905.06014 [math-ph]}}.

\bibitem{Dri87}

V.~G. Drinfeld, \emph{Quantum groups}, Proceedings of the International
  Congress of Mathematicians, Berkeley, 1986 (A.~E. Gleason, ed.), vol.~1,
  American Mathematical Society, Providence, 1987, pp.~798--820.

\bibitem{Jim85}

M.~Jimbo, \emph{A $q$-difference analogue of {$\mathrm U(\mathfrak g)$} and the
  {Y}ang-{B}axter equation}, \href{http://dx.doi.org/10.1007/BF00704588}{Lett.
  Math. Phys.} \textbf{10} (1985), 63--69.

\bibitem{BooGoeKluNirRaz13}

H.~Boos, F.~G{\"o}hmann, A.~Kl{\"u}mper, Kh.~S. Nirov, and A.~V. Razumov,
  \emph{Universal integrability objects},
  \href{http://dx.doi.org/10.1007/s11232-013-0002-8}{Theor. Math. Phys.}
  \textbf{174} (2013), 21--39, \href{http://arxiv.org/abs/1205.4399}{{\tt
  arXiv:1205.4399 [math-ph]}}.

\bibitem{BooGoeKluNirRaz14a}

H.~Boos, F.~G{\"o}hmann, A.~Kl\"umper, Kh.~S. Nirov, and A.~V. Razumov,
  \emph{Universal ${R}$-matrix and functional relations},
  \href{http://dx.doi.org/10.1142/S0129055X14300052}{Rev. Math. Phys.}
  \textbf{26} (2014), 1430005 (66pp),
  \href{http://arxiv.org/abs/1205.1631}{{\tt arXiv:1205.1631 [math-ph]}}.

\bibitem{BazLukZam96}

V.~V. Bazhanov, S.~L. Lukyanov, and A.~B. Zamolodchikov, \emph{Integrable
  structure of conformal field theory, quantum {K}d{V} theory and thermodynamic
  {B}ethe ansatz}, \href{http://dx.doi.org/10.1007/BF02101898}{Commun. Math.
  Phys.} \textbf{177} (1996), 381--398,
  \href{http://arxiv.org/abs/hep-th/9412229}{{\tt arXiv:hep-th/9412229}}.

\bibitem{BazLukZam97}

V.~V. Bazhanov, S.~L. Lukyanov, and A.~B. Zamolodchikov, \emph{Integrable
  structure of conformal field theory {II}. {Q}-operator and {DDV} equation},
  \href{http://dx.doi.org/10.1007/s002200050240}{Commun. Math. Phys.}
  \textbf{190} (1997), 247--278,
  \href{http://arxiv.org/abs/hep-th/9604044}{{\tt arXiv:hep-th/9604044}}.

\bibitem{BazLukZam99}

V.~V. Bazhanov, S.~L. Lukyanov, and A.~B. Zamolodchikov, \emph{Integrable
  structure of conformal field theory {III}. {T}he {Y}ang--{B}axter relation},
  \href{http://dx.doi.org/10.1007/s002200050531}{Commun. Math. Phys.}
  \textbf{200} (1999), 297--324,
  \href{http://arxiv.org/abs/hep-th/9805008}{{\tt arXiv:hep-th/9805008}}.

\bibitem{KhoTol92}

S.~M. Khoroshkin and V.~N. Tolstoy, \emph{The uniqueness theorem for the
  universal {$R$}-matrix}, \href{http://dx.doi.org/10.1007/BF00402899}{Lett.
  Math. Phys.} \textbf{24} (1992), 231--244.

\bibitem{LevSoiStu93}

S.~Levendorskii, Ya. Soibelman, and V.~Stukopin, \emph{The quantum {W}eyl group
  and the universal quantum {$R$}-matrix for affine {L}ie algebra
  {$A_1^{(1)}$}}, \href{http://dx.doi.org/10.1007/BF00777372}{Lett. Math.
  Phys.} \textbf{27} (1993), 253--264.

\bibitem{ZhaGou94}

Y.-Z. Zhang and M.~D. Gould, \emph{Quantum affine algebras and universal
  ${R}$-matrix with spectral parameter},
  \href{http://dx.doi.org/10.1007/BF00750144}{Lett. Math. Phys.} \textbf{31}
  (1994), 101--110, \href{http://arxiv.org/abs/hep-th/9307007}{{\tt
  arXiv:hep-th/9307007}}.

\bibitem{BraGouZhaDel94}

A.~J. Bracken, M.~D. Gould, Y.-Z. Zhang, and G.~W. Delius, \emph{Infinite
  families of gauge-equivalent {$R$}-matrices and gradations of quantized
  affine algebras}, \href{http://dx.doi.org/10.1142/S0217979294001585}{Int. J.
  Mod. Phys. B} \textbf{8} (1994), 3679--3691,
  \href{http://arxiv.org/abs/hep-th/9310183}{{\tt arXiv:hep-th/9310183}}.

\bibitem{BraGouZha95}

A.~J. Bracken, M.~D. Gould, and Y.-Z. Zhang, \emph{Quantised affine algebras
  and parameter-dependent {$R$}-matrices},
  \href{http://dx.doi.org/10.1017/S0004972700014040}{Bull. Austral. Math. Soc.}
  \textbf{51} (1995), 177--194.

\bibitem{BazTsu08}

V.~V. Bazhanov and Z.~Tsuboi, \emph{Baxter's {Q}-operators for supersymmetric
  spin chains}, \href{http://dx.doi.org/10.1016/j.nuclphysb.2008.06.025}{Nucl.
  Phys. B} \textbf{805} (2008), 451--516,
  \href{http://arxiv.org/abs/0805.4274}{{\tt arXiv:0805.4274 [hep-th]}}.

\bibitem{BooGoeKluNirRaz10}

H.~Boos, F.~G{\"o}hmann, A.~Kl{\"u}mper, Kh.~S. Nirov, and A.~V. Razumov,
  \emph{Exercises with the universal {$R$}-matrix},
  \href{http://dx.doi.org/10.1088/1751-8113/43/41/415208}{J. Phys. A: Math.
  Theor.} \textbf{43} (2010), 415208 (35pp),
  \href{http://arxiv.org/abs/1004.5342}{{\tt arXiv:1004.5342 [math-ph]}}.

\bibitem{BooGoeKluNirRaz11}

H.~Boos, F.~G{\"o}hmann, A.~Kl{\"u}mper, Kh.~S. Nirov, and A.~V. Razumov,
  \emph{On the universal ${R}$-matrix for the {I}zergin--{K}orepin model},
  \href{http://dx.doi.org/10.1088/1751-8113/44/35/355202}{J. Phys. A: Math.
  Theor.} \textbf{44} (2011), 355202 (25pp),
  \href{http://arxiv.org/abs/1104.5696}{{\tt arXiv:1104.5696 [math-ph]}}.

\bibitem{Raz13}

A.~V. Razumov, \emph{Monodromy operators for higher rank},
  \href{http://dx.doi.org/10.1088/1751-8113/46/38/385201}{J. Phys. A: Math.
  Theor.} \textbf{46} (2013), 385201 (24pp),
  \href{http://arxiv.org/abs/1211.3590}{{\tt arXiv:1211.3590 [math.QA]}}.

\bibitem{BazHibKho02}

V.~V. Bazhanov, A.~N. Hibberd, and S.~M. Khoroshkin, \emph{Integrable structure
  of {$\mathcal W_3$} conformal field theory, quantum {B}oussinesq theory and
  boundary affine {T}oda theory},
  \href{http://dx.doi.org/10.1016/S0550-3213(01)00595-8}{Nucl. Phys. B}
  \textbf{622} (2002), 475--574,
  \href{http://arxiv.org/abs/hep-th/0105177}{{\tt arXiv:hep-th/0105177}}.

\bibitem{Koj08}

T.~Kojima, \emph{Baxter's ${Q}$-operator for the ${W}$-algebra ${W_N}$},
  \href{http://dx.doi.org/10.1088/1751-8113/41/35/355206}{J. Phys. A: Math.
  Theor} \textbf{41} (2008), 355206 (16pp),
  \href{http://arxiv.org/abs/0803.3505}{{\tt arXiv:0803.3505 [nlin.SI]}}.

\bibitem{BooGoeKluNirRaz14b}

H.~Boos, F.~G{\"o}hmann, A.~Kl\"umper, Kh.~S. Nirov, and A.~V. Razumov,
  \emph{Quantum groups and functional relations for higher rank},
  \href{http://dx.doi.org/10.1088/1751-8113/47/27/275201}{J. Phys. A: Math.
  Theor.} \textbf{47} (2014), 275201 (47pp),
  \href{http://arxiv.org/abs/1312.2484}{{\tt arXiv:1312.2484 [math-ph]}}.

\bibitem{NirRaz14}

Kh.~S. Nirov and A.~V. Razumov, \emph{Quantum groups and functional relations
  for lower rank}, \href{http://arxiv.org/abs/1412.7342}{{\tt arXiv:1412.7342
  [math-ph]}}.

\bibitem{JimMiw95}

M.~Jimbo and T.~Miwa, \emph{Algebraic analysis of solvable lattice models},
  Regional Conference Series in Mathematics, no.~85, American Mathematical
  Society, Providence, 1995.

\bibitem{FreRes92}

I.~B. Frenkel and N.~Yu. Reshetikhin, \emph{Quantum affine algebras and
  holonomic difference equations},
  \href{http://dx.doi.org/10.1007/BF02099206}{Commun. Math. Phys.} \textbf{146}
  (1992), 1--60.

\bibitem{NirRaz19}

Kh.~S. Nirov and A.~V. Razumov, \emph{Vertex models and spin chains in formulas
  and pictures}, \href{http://dx.doi.org/10.3842/SIGMA.2019.068}{SIGMA} (2019),
  no.~15, 068 (67pp), \href{http://arxiv.org/abs/1811.09401}{{\tt
  arXiv:1811.09401 [math-ph]}}.

\bibitem{Ser01}

J.-P. Serre, \emph{Complex semisimple {Lie} algebras}, Springer Monographs in
  Mathematics, Springer, Berlin, 2001.

\bibitem{Hum80}

J.~E. Humphreys, \emph{Introduction to {L}ie algebras and representation
  theory}, Springer, New York, 1980.

\bibitem{Kac90}

V.~Kac, \emph{Infinite dimensional {L}ie algebras}, Cambridge University Press,
  Cambridge, 1990.

\bibitem{Tan92}

T.~Tanisaki, \emph{{K}illing forms, {H}arish-{C}handra homomorphisms and
  universal ${R}$-matrices for quantum algebras}, Infinite Analysis
  (A.~Tsuchiya, T.~Eguchi, and M.~Jimbo, eds.), Advanced Series in Mathematical
  Physics, vol.~16, World Scientific, Singapore, 1992, pp.~941--962.

\bibitem{FreRes99}

E.~Frenkel and N.~Reshetikhin, \emph{The $q$-characters of representations of
  quantum affine algebras and deformations of ${W}$-algebras}, Contemp. Math.
  \textbf{248} (1999), 163--205, \href{http://arxiv.org/abs/math/9810055}{{\tt
  arXiv:math/9810055}}.

\bibitem{MukYou14}

E.~Mukhin and C.~A.~S. Young, \emph{Affinization of category $\mathcal{O}$ for
  quantum groups},
  \href{http://dx.doi.org/10.1090/S0002-9947-2014-06039-X}{Trans. Amer. Math.
  Soc.} \textbf{366} (2014), 4815--4847,
  \href{http://arxiv.org/abs/1204.2769}{{\tt arXiv:1204.2769 [math.QA]}}.

\bibitem{EtiFreKir98}

P.~Etingof, B.~Frenkel, and A.~A. Kirillov, \emph{Lectures on representation
  theory and {K}nizhnik--{Z}amolodchikov equations}, Mathematical Surveys and
  Monographs, vol.~58, American Mathematical Society, Providence, 1998.

\bibitem{BooJimMiwSmiTak07}

H.~Boos, M.~Jimbo, T.~Miwa, F.~Smirnov, and Y.~Takeyama, \emph{Hidden
  {G}rassmann structure in the {XXZ} model},
  \href{http://dx.doi.org/10.1007/s00220-007-0202-x}{Commun. Math. Phys.}
  \textbf{272} (2007), 263--281,
  \href{http://arxiv.org/abs/hep-th/0606280}{{\tt arXiv:hep-th/0606280}}.

\bibitem{BooJimMiwSmiTak09}

H.~Boos, M.~Jimbo, T.~Miwa, F.~Smirnov, and Y.~Takeyama, \emph{Hidden
  {G}rassmann structure in the {XXZ} model {II}: {C}reation operators},
  \href{http://dx.doi.org/10.1007/s00220-008-0617-z}{Commun. Math. Phys.}
  \textbf{286} (2009), 875--932, \href{http://arxiv.org/abs/0801.1176}{{\tt
  arXiv:0801.1176 [hep-th]}}.

\bibitem{BooJimMiwSmiTak06b}

H.~Boos, M.~Jimbo, T.~Miwa, F.~Smirnov, and Y.~Takeyama, \emph{Reduced q{KZ}
  equation and correlation functions of the {XXZ} model},
  \href{http://dx.doi.org/10.1007/s00220-005-1430-6}{Commun.Math.Phys.}
  \textbf{261} (2006), 245--276,
  \href{http://arxiv.org/abs/hep-th/0412191}{{\tt arXiv:hep-th/0412191}}.

\bibitem{Jim86a}

M.~Jimbo, \emph{A $q$-analogue of {$\mathrm U(\mathfrak{gl}(N + 1))$}, {H}ecke
  algebra, and the {Y}ang--{B}axter equation},
  \href{http://dx.doi.org/10.1007/BF00400222}{Lett. Math. Phys.} \textbf{11}
  (1986), 247--252.

\bibitem{IdzTokIohJimMiw93}

M.~Idzumi, T.~Tokihiro, K.~Iohara, M.~Jimbo, and T.~Miwa, \emph{Quantum affine
  symmetry in vertex models},
  \href{http://dx.doi.org/10.1142/S0217751X9300062X}{Int. J. Mod. Phys. A}
  \textbf{8} (1993), 1479--1511,
  \href{http://arxiv.org/abs/hep-th/9208066}{{\tt arXiv:hep-th/9208066}}.

\bibitem{Koj18}

T.~Kojima, \emph{Commutation relations of vertex operators for
  ${U_q(\widehat{sl}(M | N))}$}, \href{http://dx.doi.org/10.1063/1.5047255}{J.
  Math. Phys.} \textbf{59} (2018), 101701,
  \href{http://arxiv.org/abs/1807.02273}{{\tt arXiv:1807.02273 [math.QA]}}.

\bibitem{Auf11}

B.~Aufgebauer, \emph{Berechnung der {K}orrelationsfunktionen des
  {H}eisenberg-{M}odells bei endlicher {T}emperatur mittels
  {F}unktionalgleichungen}, Ph.D. thesis, Bergische Universit\"at Wuppertal,
  2011.

\bibitem{AufKlu12}

B.~Aufgebauer and A.~Kl\"umper, \emph{Finite temperature correlation functions
  from discrete functional equations},
  \href{http://dx.doi.org/10.1088/1751-8113/45/34/345203}{J. Phys. A: Math.
  Theor.} \textbf{45} (2012), 345203 (20pp),
  \href{http://arxiv.org/abs/1205.5702}{{\tt arXiv:1205.5702
  [cond-math.stat-mech]}}.

\bibitem{BooHutNir18}

H.~Boos, A.~Hutsalyuk, and Kh.~S. Nirov, \emph{On the calculation of the
  correlation functions of the $\mathfrak{sl}_3$-model by means of the reduced
  q{KZ} equation}, \href{http://dx.doi.org/10.1088/1751-8121/aae1d6}{J. Phys.
  A: Math. Theor.} \textbf{51} (2018), 445202 (29pp),
  \href{http://arxiv.org/abs/1804.09756}{{\tt arXiv:1804.09756 [hep-th]}}.

\bibitem{RibKlu19}

G.~A.~P. Ribeiro and A.~Kl\"umper, \emph{Correlation functions of the
  integrable ${SU}(n)$ spin chain},
  \href{http://dx.doi.org/10.1088/1742-5468/aaf31e}{J. Stat. Mech.: Theor.
  Exp.} (2019), 013103 (31pp), \href{http://arxiv.org/abs/1804.10169}{{\tt
  arXiv:1804.10169 [math-ph]}}.

\bibitem{Hum96}

J.~F. Humphreys, \emph{A course in group theory}, Oxford University Press, New
  York, 1996.

\end{thebibliography}
\end{document}